\documentclass[fleqn,usenatbib,]{mnras}
\usepackage{lineno}
\usepackage{newtxtext,newtxmath}

\usepackage[T1]{fontenc}
\usepackage{ae,aecompl}

\usepackage{graphicx}	
\usepackage{amsmath}	
\usepackage{amssymb}	
\usepackage{booktabs}   
\usepackage{longtable}  
\usepackage{rotating}
\usepackage{appendix}
\usepackage{float}
\usepackage[dvipsnames]{xcolor}
\usepackage{caption}
\usepackage{multicol}
\usepackage{ctable}
\PassOptionsToPackage{hyphens}{url}\usepackage{hyperref}
\usepackage{threeparttable}

\usepackage{booktabs}

\newcommand{\halpha}[0]{H$\alpha$}
\newcommand{\hbeta}[0]{H$\beta$}
\newcommand{\hgamma}[0]{H$\gamma$}
\newcommand{\hdelta}[0]{H$\delta$}
\newcommand{\hepsilon}[0]{H$\epsilon$}
\newcommand{\HII}[0]{$[\textnormal{\textsc{Hii}}]$}
\newcommand{\OII}[0]{$[\textnormal{\textsc{Oii}}]$}
\newcommand{\OI}[0]{$[\textnormal{\textsc{Oi}}]$}
\newcommand{\OIII}[0]{$[\textnormal{\textsc{Oiii}}]$}
\newcommand{\SII}[0]{$[\textnormal{\textsc{Sii}}]$}

\newcommand{\NII}[0]{$[\textnormal{\textsc{Nii}}]$}

\usepackage{etoolbox}
\makeatletter
\patchcmd\@combinedblfloats{\box\@outputbox}{\unvbox\@outputbox}{}{%
  \errmessage{\noexpand\@combinedblfloats could not be patched}%
}%
\makeatother

\defcitealias{Pursiainen2018}{P18}
\defcitealias{Wiseman2020}{W20}
\defcitealias{Pursiainen2020}{P20}

\title[RET host galaxies in DES]{The Host Galaxies of Rapidly Evolving Transients in the Dark Energy Survey}
\author[P. S. Wiseman]{
\parbox{\textwidth}{
\Large
P.~Wiseman,$^{1}$
M.~Pursiainen,$^{1}$
M.~Childress,$^{1}$
E.~Swann,$^{2}$
M.~Smith,$^{1}$
L.~Galbany,$^{3}$
C.~Lidman,$^{4}$
T.~M.~Davis,$^{5}$
C.~P.~Guti\'errez,$^{1}$
A.~M\"oller,$^{6}$
B.~P.~Thomas,$^{2}$
C.~Frohmaier,$^{2}$
R.~J.~Foley,$^{7}$
S.~R.~Hinton,$^{5}$
L.~Kelsey,$^{1}$
R.~Kessler,$^{8,9}$
G.~F.~Lewis,$^{10}$
M.~Sako,$^{11}$
D.~Scolnic,$^{12}$
M.~Sullivan,$^{1}$
M.~Vincenzi,$^{2}$
T.~M.~C.~Abbott,$^{13}$
M.~Aguena,$^{14,15}$
S.~Allam,$^{16}$
J.~Annis,$^{16}$
E.~Bertin,$^{17,18}$
S.~Bhargava,$^{19}$
D.~Brooks,$^{20}$
D.~L.~Burke,$^{21,22}$
A.~Carnero~Rosell,$^{23}$
D.~Carollo,$^{24}$
M.~Carrasco~Kind,$^{25,26}$
J.~Carretero,$^{27}$
M.~Costanzi,$^{28,29}$
L.~N.~da Costa,$^{15,30}$
H.~T.~Diehl,$^{16}$
P.~Doel,$^{20}$
S.~Everett,$^{7}$
P.~Fosalba,$^{31,32}$
J.~Frieman,$^{16,9}$
J.~Garc\'ia-Bellido,$^{33}$
E.~Gaztanaga,$^{31,32}$
K.~Glazebrook,$^{34}$
D.~Gruen,$^{35,21,22}$
R.~A.~Gruendl,$^{25,26}$
J.~Gschwend,$^{15,30}$
G.~Gutierrez,$^{16}$
D.~L.~Hollowood,$^{7}$
K.~Honscheid,$^{36,37}$
D.~J.~James,$^{38}$
K.~Kuehn,$^{39,40}$
N.~Kuropatkin,$^{16}$
M.~Lima,$^{14,15}$
M.~A.~G.~Maia,$^{15,30}$
J.~L.~Marshall,$^{41}$
P.~Martini,$^{36,42}$
F.~Menanteau,$^{25,26}$
R.~Miquel,$^{43,27}$
A.~Palmese,$^{16,9}$
F.~Paz-Chinch\'{o}n,$^{44,26}$
A.~A.~Plazas,$^{45}$
A.~K.~Romer,$^{19}$
E.~Sanchez,$^{23}$
V.~Scarpine,$^{16}$
M.~Schubnell,$^{46}$
S.~Serrano,$^{31,32}$
I.~Sevilla-Noarbe,$^{23}$
N.~E.~Sommer,$^{4}$
E.~Suchyta,$^{47}$
M.~E.~C.~Swanson,$^{26}$
G.~Tarle,$^{46}$
B.~E.~Tucker,$^{4}$
D.~L.~Tucker,$^{16}$
T.~N.~Varga,$^{48,49}$
and A.~R.~Walker$^{13}$
\begin{center} (DES Collaboration) \end{center}
}
\vspace{0.4cm}
\\
\parbox{\textwidth}{Affiliations are listed at the end of the paper}
}
\date{Accepted XXX. Received YYY; in original form ZZZ}

\pubyear{2020}

\begin{document}
\label{firstpage}
\pagerange{\pageref{firstpage}--\pageref{lastpage}}
\maketitle

\begin{abstract}
Rapidly evolving transients (RETs), also termed fast blue optical transients (FBOTs), are a distinct class of astrophysical event. They are characterised by lightcurves that decline much faster than common classes supernovae (SNe), span vast ranges in peak luminosity and can be seen to redshifts greater than 1. Their evolution on fast timescales has hindered high quality follow-up observations, and thus their origin and explosion/emission mechanism remains unexplained. In this paper we define the largest sample of RETs to date, comprising 106 objects from the Dark Energy Survey, and perform the most comprehensive analysis of RET host galaxies. Using deep-stacked photometry and emission-lines from OzDES spectroscopy, we derive stellar masses and star-formation rates (SFRs) for 49 host galaxies, and metallicities for 37. We find that RETs explode exclusively in star-forming galaxies and are thus likely associated with massive stars. Comparing RET hosts to samples of host galaxies of other explosive transients as well as field galaxies, we find that RETs prefer galaxies with high specific SFRs, indicating a link to young stellar populations, similar to stripped-envelope SNe. RET hosts appear to show a lack of chemical enrichment, their metallicities akin to long duration gamma-ray bursts and superluminous SN host galaxies. There are no clear relationships between properties of the host galaxies and the peak magnitudes or decline rates of the transients themselves.

\end{abstract}




\section{Introduction}

In the standard paradigm of stellar evolution, stars with a zero-age main sequence (ZAMS) mass above $8M_{\sun}$ are believed to explode as a result of a catastrophic collapse of their iron cores and are known as core-collapse supernovae (CCSNe). CCSNe can be split into observationally-determined subclasses based on their lightcurve and spectral evolution: SNe II display hydrogen features in their spectra \citep{Minkowski1941}, and are thought to occur in stars that retain a large fraction of their hydrogen envelope. Conversely, SNe Ib and Ic do not show signatures of hydrogen \citep[e.g.][]{Filippenko1997} and are thus referred to collectively as stripped-envelope SNe (SESNe), having undergone a partial removal of their outer atmospheres. The SN IIb subclass, which shows hydrogen only at early epochs that disappears after a few weeks \citep{Filippenko1988}, is also commonly grouped along with SESNe. SNe IIn display much narrower hydrogen emission lines when compared to standard SNe II \citep{Schlegel1990}. The narrow emission originates from the ejecta impacting on slow-moving circumstellar material (CSM). Since the turn of the century, observations of CCSNe, whose lightcurves are primarily powered by the radioactive decay of freshly synthesised Ni-56, have been supplemented by rarer, more exotic transient classes.

 Long duration gamma-ray bursts (LGRBs), although first discovered in the 1960s \citep{Klebesadel1973}, were unequivocally linked to collapsing massive stars through their associations with broad-lined type Ic SNe \citep{Galama1998,Hjorth2003}. Thought to be caused by accretion onto a newly-formed black hole at the centre of a collapsing, rapidly-rotating massive star \citep[e.g.][]{Woosley1993,Woosley2006a,Woosley2006b}, LGRBs comprise roughly $1\%$ of all SNe Ic, which themselves make up only 15\% of all CCSNe \citep{Kelly2012,Graham2016}. Another exotic class of SNe is the particularly bright superluminous supernovae (SLSNe; e.g. \citealt{Quimby2011, Gal-Yam2012}). Originally grouped due to their slowly-evolving lightcurves and extreme luminosity (peaking at $M_B < -21$~mag; 10-100 times brighter than typical CCSNe), recent observations have revealed a continuum of spectroscopically similar objects with peaks as faint as $M_B \sim -19$~mag \citep{DeCia2018,Lunnan2018,Angus2019}, which overlaps with the bright end of the CCSN luminosity function \citep{Li2011}. The lightcurve evolution of SLSNe is not well described by models of Ni-56 decay, with the most popular alternative hypothesis being the magnetic coupling of the ejecta with the spin down of a newly formed, rapidly rotating magnetar.
 
 Along with observations of the transients themselves, studies of host galaxies are frequently used to make strong inferences about the progenitor stars and explosion mechanisms. CCSNe are confined almost exclusively to galaxies hosting recent or ongoing star formation, due to their origin from massive stars. There are correlations between the expected progenitor mass of different sub-classes of CCSNe and host galaxy properties. On average, SESNe reside in galaxies with higher specific star-formation rates (sSFRs; \citealt{James2006,Kelly2008}), while studies of the local environments tend to show that SESNe explode closer to \HII~regions than SNe II, indicating that the progenitors are younger and more massive than the various sub-classes of hydrogen-rich SNe II  \citep[e.g.][]{Anderson2012,Galbany2018}. More extreme events tend to occur in galaxies low in mass and high in sSFR, with both LGRBs \citep[e.g.][]{Fruchter2006,LeFloch2006,Levesque2010,Kruehler2015,Vergani2015,Perley2016b,Palmerio2019,Taggart2019} and to an even greater degree SLSNe \citep[e.g.][]{Neill2011,Lunnan2014,Leloudas2015,Angus2016,Schulze2018,Taggart2019} exhibiting this association.
 
 The chemical composition of the interstellar medium (ISM) is an important consideration when comparing host galaxy properties. While it does not appear to play a significant role in the relative production of CCSNe (although there are some trends, with SESNe typically found in slightly less metal-rich environments than SNe II; \citealt{Galbany2018}), it appears to be vitally important in the production of LGRBs and SLSNe. Theory predicts that the production of a LGRB should only be possible in stars with a metallicity of $Z/Z_{\sun}\leq 0.3$ \citep{Woosley1993}  in order for the likely Wolf-Rayet or blue supergiant progenitors not to lose their outer atmospheres through metal-driven winds, thus conserving sufficient angular momentum to power the black-hole-driven jet or rapidly rotating magnetar. Many LGRB host galaxy studies have indeed revealed a metallicity threshold between 0.5 and 1 times the solar value \citep[e.g.][]{Stanek2006,Modjaz2008,Kruehler2015,Perley2016b,Japelj2016,Vergani2017}, although this is not a rigid threshold.  
SLSN host galaxies also appear to be lower in metallicity than would be expected for their stellar mass, with a suppression of SLSN production above a value of about half-solar \citep{Lunnan2014,Chen2016a,Perley2016c}. Like LGRBs, SLSNe also require a particularly high sSFR, suggesting that they are explosions of very young, rapidly rotating massive stars.
 
Recently, inspection of high-cadence, large-area survey data sets have revealed more exotic transients that are difficult to explain with conventional models. \citet{Drout2014} presented a sample of rapidly evolving transients (RETs; also termed `Fast Blue Optical Transients' - FBOTs or `Fast Evolving Luminous Transients' - FELTs) in the Pan-STARRS survey (PS1). \citet{Arcavi2016} reported three rapidly rising, highly luminous objects in the Supernova Legacy Survey (SNLS) and one in the Palomar Transient Factory (PTF), of which one (SNLS04D4ec) declines rapidly like the PS1 sample. \citet[hereafter P18]{Pursiainen2018} expanded the known number of RETs to beyond 80 with their sample from the Dark Energy Survey (DES), spanning a redshift range of $\sim 0$ to $>1$. A further sample of five objects has been discovered by the Hyper Suprime-Cam Subaru Strategic Program (SSP) Transient Survey \citep{Tampo2020}. RETs typically rise to peak brightness in less than 10 days, and decline to 10\% of their peak brightness within 30 days, much faster than typical SNe. The photometric measurements of the PS1 and DES RETs seem to be well described by expanding blackbodies, although a handful show declining photospheric radii from the first detection. Due to the rapid nature of their lightcurves and location at high-redshift, spectral coverage is sparse and signal-to-noise ratio (SNR) is low, such that there has not yet been a conclusive detection of absorption or emission features from the transients and thus the physical mechanism responsible for their rapid evolution remains unexplained.

There are a limited number of events detected in the local Universe whose properties appear consistent with the RETs seen at cosmological distances in the PS1 and DES samples, the most widely studied of which is AT2018cow \citep[e.g.][]{Prentice2018,Margutti2019,Perley2019}. The transient declined from its discovery, with constraints on the rise time of 1 day, and from X-rays through to radio wavelengths did not resemble any known SN, GRB afterglow, or kilonova (KN; \citealt{Ho2019}) . There are many diverse explanations for the power source of AT2018cow touted in the literature, including: magnetars \citep{Mohan2020}; electron capture collapse of merged white dwarfs \citep{Lyutikov2019}; a tidal disruption event (TDE) of a white dwarf \citep{Kuin2019} or of a main sequence star by an intermediate mass black hole \citep{Perley2019}; common envelope jets supernova (CEJSN; \citealt{Soker2019}); or a wind-driven transient \citep{Uno2020}.
Other nearby rapid transients include the local fast-declining SN-like transient SN2018kzr \citep{McBrien2019} which is explained by the accretion-induced collapse of a white dwarf or a white dwarf-neutron star merger, and KSN-2015K \citep{Rest2018} whose fast rise and decline is explained by the shock of an SN running into previously-expelled material.  It is currently unclear whether these transients represent the local analogues of the DES and PS1 RETs.

In this paper, we present the first comprehensive study of the host galaxies of RETs. We make use of the final DES sample, which builds on \citetalias{Pursiainen2018} by adding the 5th and final season of DES-SN observations as well as more refined selection techniques. Using the deep DES photometry from \citet[hereafter W20]{Wiseman2020} and spectra from OzDES \citep{Lidman2020} we derive host galaxy properties and compare them to samples of host galaxies of CCSNe, LGRBs, and SLSNe, as well as the individual local rapid transients. For clarity, we will use the term RET to refer only to events in the high-redshift samples of DES and PS1. 

The order of the paper is as follows: in Section \ref{sec:sample} we introduce the full DES RET sample and describe the host galaxy observations in Section \ref{sec:obs}. The analysis methods and results are described in Sections \ref{sec:measure} and \ref{sec:analysis} respectively, before a discussion (Section \ref{sec:disc}) and conclusion (Section \ref{sec:conc}).
Where applicable, we adopt a spatially flat $\Lambda$CDM cosmology with the parameters $H_0=70$ km~s$^{-1}$ Mpc$^{-1}$ and $\Omega_{\textrm{M}}=0.3$. Magnitudes are presented in the AB system \citep{Oke1983}, and values (uncertainties) are quoted at the 50\textsuperscript{th} (16\textsuperscript{th} and 84\textsuperscript{th}, i.e. $1\sigma$) percentiles of their probability density function.

\begin{figure}
\includegraphics[width=0.5\textwidth]{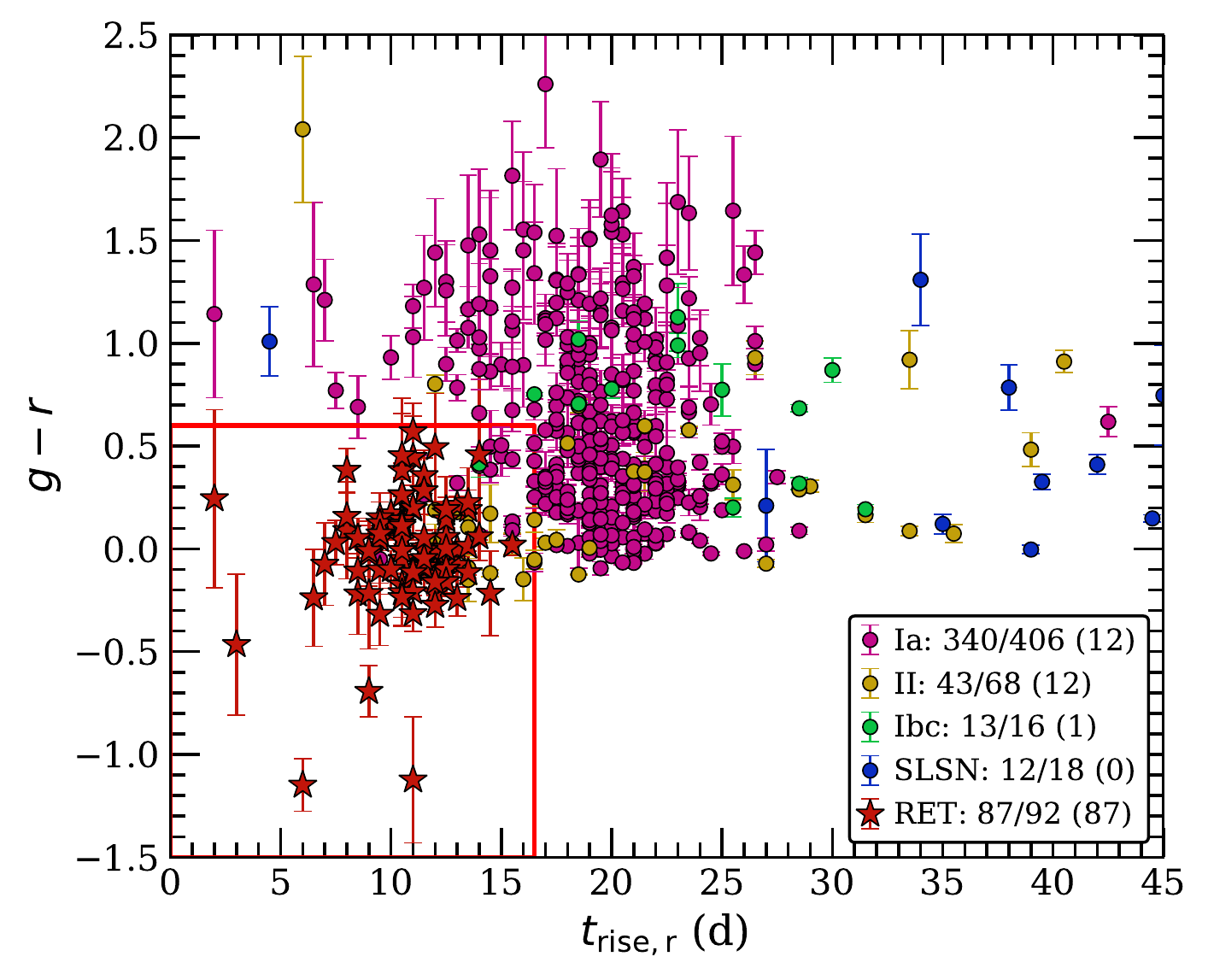}
\caption{Observer-frame $g-r$ colour at maximum light and observer-frame $r$-band rise-time, derived from Gaussian Processed fits to DES-SN photometry for spectroscopically confirmed SNe and \citetalias{Pursiainen2018} RETs. The location of the red box is designed to maximise the completeness and purity of RETs. The fractions in the legend refer to the number of each class of transient passing the cuts of Section \ref{subsec:new_method} compared to the total number of that class in DES-SN, while the number in parentheses refers to the number inside the red box defining RET parameter space. 
\label{fig:selection}}
\end{figure}

\begin{table*}
\caption{Host galaxy information for the 106 RETs in the DES 5-year sample. A full-length, machine-readable version of this table is available in the online version of the manuscript.}

\begin{threeparttable}
\begin{tabular}{llllllll}

\toprule
{} &       RA &       Dec &    $m_r$\tnote{a}  &     $z$ &     Survey & Exposure time \tnote{b} \\
Transient name &  (J2000  & (J2000)  &  (AB)    &         &         &   (hours)         &               \\
\midrule
DES13C1acmt    & 54.32925 & -26.83371 & $23.18 \pm 0.04$ &    $^c$ &      OzDES &             - \\
DES13C1tgd     & 54.06436 & -27.63867 & $20.30 \pm 0.05$ & 0.19647 &      OzDES &      10.17 \\
DES13C3abtt    & 52.62108 & -28.16151 & $21.50 \pm 0.03$ &    $^d$ &          - &             - \\
DES13C3asvu    & 52.83670 & -27.36071 & $22.46 \pm 0.02$ &    $^d$ &          - &             - \\
DES13C3avkj    & 51.97076 & -27.52792 & $24.07 \pm 0.04$ &    $^c$ &      OzDES &             - \\
DES13C3bcok    & 53.02711 & -28.62476 & $18.61 \pm 0.02$ & 0.34577 &  LADUMA &               \\
DES13C3nxi     & 51.96356 & -28.35720 & $24.99 \pm 0.05$ &    $^c$ &      OzDES &             - \\
DES13C3smn     & 51.97112 & -28.08362 & $25.30 \pm 0.06$ &    $^c$ &      OzDES &             - \\
DES13C3uig     & 52.94416 & -27.58544 & $22.17 \pm 0.03$ & 0.67346 &       ACES &               \\
DES13E2lpk     & 10.09911 & -43.53903 & $20.49 \pm 0.03$ & 0.47541 &      OzDES &       6.00 \\
DES13S2wxf\tnote{e}& 41.61268 & -0.02634&$21.39 \pm0.03$9& 0.56985 &      OzDES &       5.42 \\
DES13X1hav     & 35.03245 &  -5.11022 & $23.63 \pm 0.07$ & 0.58236 &      OzDES &       4.00 \\
DES13X2oyb     & 35.31056 &  -5.67893 & $22.91 \pm 0.06$ &    $^c$ &      OzDES &             - \\
DES13X2wvv\tnote{e}     & 34.86526 &  -6.71603 & $21.89 \pm 0.02$ & 0.47503 &      OzDES &       4.50 \\
DES13X3aakf    & 35.71205 &  -4.69915 & $25.47 \pm 0.09$ &    $^c$ &      OzDES &             - \\
DES13X3afjd    & 37.00386 &  -4.58049 & $20.75 \pm 0.05$ &    $^d$ &          - &             - \\
DES13X3alnb    & 37.18496 &  -5.14232 & $25.04 \pm 0.06$ &    $^c$ &      OzDES &             - \\
DES13X3gmd\tnote{e}     & 36.50409 &  -4.21949 & $22.90 \pm 0.05$ & 0.78082 &      OzDES &      13.92 \\
DES13X3gms     & 35.80095 &  -4.49384 & $23.02 \pm 0.06$ & 0.64792 &      OzDES &       6.58 \\
DES13X3kgm     & 36.50380 &  -4.86636 & $26.46 \pm 0.18$ &    $^d$ &          - &             - \\
DES13X3npb     & 36.64218 &  -4.13381 & $20.79 \pm 0.05$ & 0.49542 &      OzDES &       5.17 \\
DES13X3nyg     & 36.99228 &  -3.91327 & $23.38 \pm 0.06$ & 0.71205 &      OzDES &      45.50 \\
DES13X3pby     & 36.33330 &  -5.31408 & $23.82 \pm 0.04$ &    $^c$ &      OzDES &             - \\
DES14C1jnd     & 54.35153 & -27.49300 & $24.33 \pm 0.08$ &    $^c$ &      OzDES &             - \\
DES14C3gzj\tnote{e}     & 52.57168 & -28.14019 & $26.04 \pm 0.10$ &    $^d$ &          - &             - \\
DES14C3tnz\tnote{e}     & 52.86248 & -28.51318 & $22.26 \pm 0.03$ & 0.70452 &      OzDES &       5.79 \\
DES14C3tvw     & 53.32230 & -27.90621 & $21.31 \pm 0.02$ & 0.70390 &       ACES &               \\
DES14E1aqi\tnote{e}     &  8.73970 & -43.40182 & $25.61 \pm 0.23$ &    $^d$ &          - &             - \\
DES14E2xsm     &  9.67624 & -43.58736 & $23.39 \pm 0.04$ &    $^c$ &      OzDES &             - \\
DES14S2anq     & 41.27776 &  -0.74529 & $17.57 \pm 0.03$ & 0.05211 &      OzDES &       0.75 \\
DES14S2plb     & 41.85667 &  -1.61811 & $18.38 \pm 0.03$ & 0.11531 &      OzDES &       2.00 \\
DES14S2pli     & 41.22837 &  -1.09830 & $20.86 \pm 0.03$ & 0.35478 &      OzDES &       3.42 \\
DES14X1bnh     & 33.74902 &  -4.79254 & $21.95 \pm 0.03$ & 0.82982 &      OzDES &       3.50 \\
DES14X3pkl     & 37.21103 &  -4.80741 & $22.40 \pm 0.06$ & 0.29537 &      OzDES &       5.17 \\
DES14X3pko     & 36.90824 &  -3.69642 & $24.48 \pm 0.06$ &    $^d$ &          - &             - \\
DES15C2eal     & 54.06180 & -29.23000 & $22.99 \pm 0.03$ & 0.22347 &      OzDES &      16.91 \\
DES15C3edw     & 52.55235 & -27.71022 & $22.69 \pm 0.03$ &    $^d$ &          - &             - \\
DES15C3lpq     & 52.71204 & -28.61319 & $23.28 \pm 0.03$ & 0.61365 &      OzDES &       6.17 \\
DES15C3lzm     & 52.17460 & -28.23186 & $20.47 \pm 0.02$ & 0.32690 &      ATLAS &               \\
DES15C3mem\tnote{e}     & 52.14216 & -29.05851 & $20.28 \pm 0.04$ & 0.61618 &     PRIMUS &               \\
DES15C3mgq     & 52.76901 & -28.20882 & $22.97 \pm 0.03$ & 0.23031 &      OzDES &       8.17 \\
DES15C3nat     & 52.88528 & -28.72370 & $23.39 \pm 0.03$ & 0.83929 &      OzDES &      18.62 \\
DES15C3opk     & 51.66147 & -28.34737 & $23.05 \pm 0.03$ & 0.56984 &      OzDES &      12.17 \\
DES15C3opp     & 51.73962 & -28.11496 & $23.25 \pm 0.04$ & 0.44242 &      OzDES &      16.75 \\
DES15C3pbi     & 52.23620 & -28.00223 & $25.10 \pm 0.05$ &    $^d$ &          - &             - \\
DES15E2lmq     &  9.62005 & -43.98734 & $25.65 \pm 0.18$ &    $^d$ &          - &             - \\
DES15E2nqh     &  9.73174 & -43.08707 & $23.29 \pm 0.03$ & 0.51525 &      OzDES &       6.83 \\
DES15S1fli     & 43.18790 &  -0.88620 & $20.98 \pm 0.02$ & 0.44739 &      OzDES &       2.33 \\
DES15S1fll     & 42.78846 &  -0.19747 & $20.95 \pm 0.03$ & 0.22647 &      OzDES &       0.67 \\
DES15X2ead     & 36.48913 &  -6.45118 & $20.08 \pm 0.03$ & 0.23175 &      OzDES &       2.00 \\
DES15X3atd     & 35.84011 &  -4.29142 & $23.87 \pm 0.05$ &    $^c$ &      OzDES &             - \\
DES15X3kyt     & 36.27500 &  -5.41103 & $24.74 \pm 0.07$ &    $^d$ &          - &             - \\

\bottomrule
\end{tabular}
\begin{tablenotes}
\item[a] Apparent $r$-band Kron magnitude according to DES-SN deep coadds of \citetalias{Wiseman2020}, not corrected for Galactic foreground reddening.
\item[b] Exposure time only given for spectra which we have used for line measurements rather than just redshift.
\item[c] Host targetted by OzDES but no redshift measurement possible.
\item[d] Host not targetted by OzDES.
\item[e] Found by updated search method (Section \ref{subsec:new_method}).
\item[f] Found in Season 5 using \citetalias{Pursiainen2018} method.
\item[g] Redshift updated since to \citetalias{Pursiainen2018}.
\end{tablenotes}
\end{threeparttable}
\label{tab:obs}
\end{table*}

\begin{figure*}
\includegraphics[width=\linewidth]{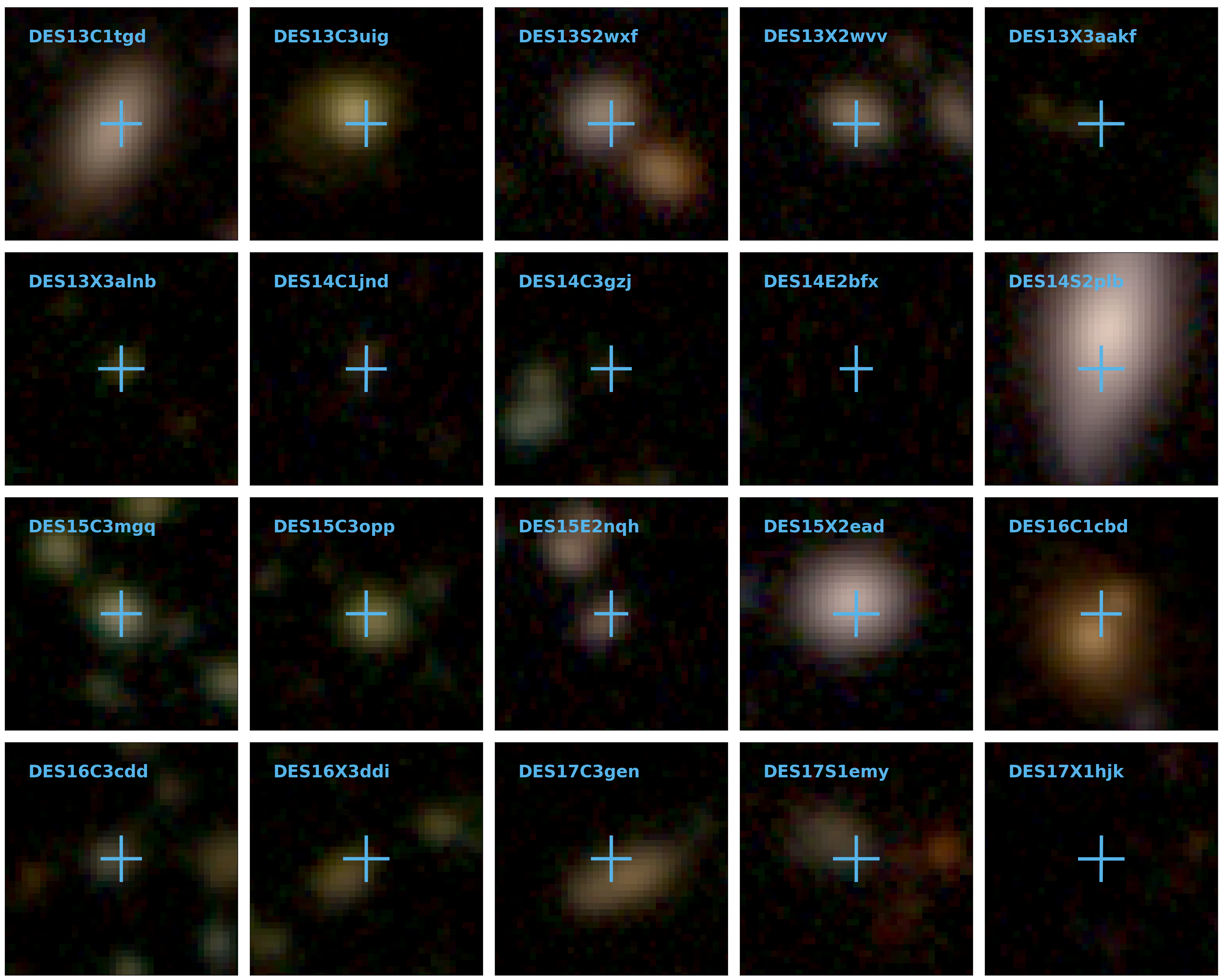}
\caption{Selection of DES RET host galaxies in an RGB composite of the DES $gri$ band deep coadds from \citetalias{Wiseman2020}. The locations of the transients are indicated with cyan crosses. The stamps have a size of 10\arcsec in each direction. DES14E2bfx and DES17X1hjk are considered hostless.
\label{fig:mosaic}}
\end{figure*}

\begin{figure}
\includegraphics[width=0.5\textwidth]{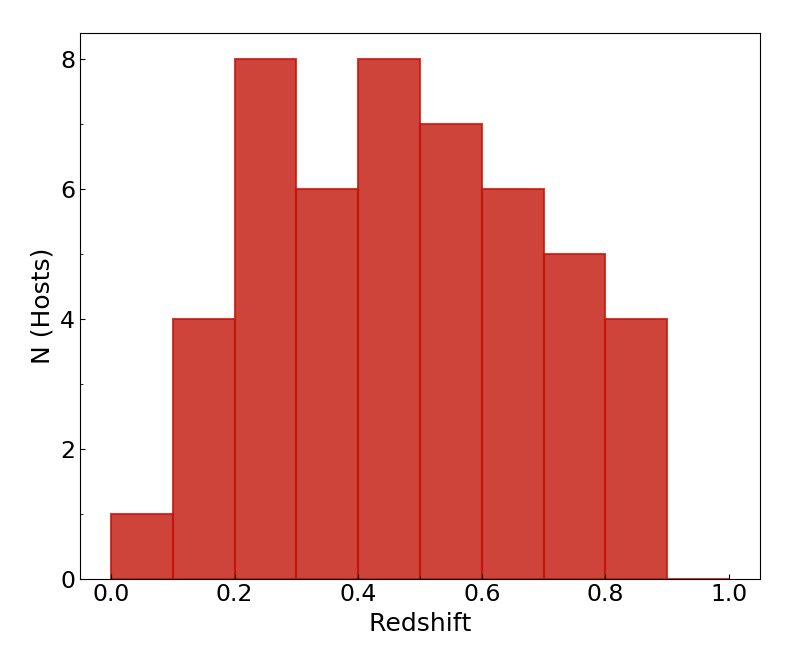}
\caption{Redshift distribution for the host galaxies of RETs in DES for which a measurement was obtained.
\label{fig:z_dist}}
\end{figure}

\section{Sample selection}
\label{sec:sample}
We derive our sample from the 106 RETs discovered in the 5-year DES-SN transient survey. This number expands upon the 72 of \citetalias{Pursiainen2018}. The first reason for the increased sample size is the use of the 5th year of DES-SN, as \citetalias{Pursiainen2018} were only able to make use of the first four years. By imposing the \citetalias{Pursiainen2018} selection criteria on season 5, the sample is increased to 92 objects. The second reason is an update to the sample selection technique, outlined in the following subsection, which adds a further 14 transients.

\subsection{The DES supernova programme \label{subsec:desdata}}

The Dark Energy Survey (DES) makes use of the Dark Energy Camera (DECam; \citealt{Flaugher2015}) to survey $5000\deg^2$ of the southern sky between 2013 and 2019. The supernova programme (DES-SN), designed primarily to detect type Ia SNe (SNe Ia) for cosmological measurements \citep{Bernstein2012}, consisted of five 6-month seasons of approximately 7 day cadence in ten single pointings, known as the SN fields which cover a total area of $\sim27 \deg^2$. Of these fields, two were observed with longer exposure times and are referred to as deep fields with remaining eight known as shallow fields; the resulting approximate single-visit depths are 23.5 mag (shallow) and 24.5 mag (deep) in the DES $r$-band. Transient detection was performed by difference imaging using a custom pipeline \citep{Kessler2015}. Transient candidates were vetted via machine-learning techniques \citep{Goldstein2015}, leading to $\sim 30,000$ viable supernova candidates over the full five seasons. DES-SN included an extensive spectroscopic follow-up program to identify transients and measure spectroscopic redshifts of host galaxies \citep{DAndrea2018}. A full description of the search for rapid transients in DES can be found in \citetalias{Pursiainen2018}.

\subsection{Improvements to the search method \label{subsec:new_method}}

The original method of finding RETs in the DES-SN data (and presented in \citetalias{Pursiainen2018}) was designed to be simple and used light curve modelling with Gaussian and linear fits. The simplistic method made it possible to look for exotic transients without knowing their observed characteristics beforehand and resulted in a large sample of photometrically selected fast transients. However, as the search was simplistic and relied heavily on visual inspection of the available data (etc. images, light curves, host galaxy information), it is impossible to quantify the completeness of the sample. For instance, due to the large redshift range within the sample it is entirely possible that distant events with longer rise times could have been missed due to time dilation stretching their lightcurves, while faster events and low redshift may evolve on timescales quicker than the survey cadence. Here, a more sophisticated search method is presented. As only a fraction of transients in DES-SN have redshift information from their host galaxies, we perform the search in the observer frame. The key features of RETs that separate them from most traditional SNe types are the fast light curve evolution (rise to peak in $\lesssim15$ days) and blue colour at peak ($g-r \lesssim 0.5$). Even though both of these quantities depend on the redshift of the transient, they still effectively distinguish the the fast events from traditional SN types. We thus attempt to select a sample based on observed rise times and colours at peak brightness.

\subsubsection{Gaussian processed lightcurves}
Using observed photometric data points directly to infer rise time and colour has several problems that can be improved. For one, measuring peak colour is problematic: DES-SN did not always observe $g$- and $r$-bands on the same or even consecutive nights in the `deep' fields, thus adding larger uncertainty in the peak colour estimate. We do not have a light curve model for RETs, and therefore  measuring a 10-15 day rise time is difficult with a one week  cadence. Rather than fitting with a physically motivated model, we instead interpolate the lightcurves of all DES-SN candidates using Gaussian Processes (GP) as presented in \citet{Pursiainen2020}. The interpolated lightcurves have a 0.5 day cadence and every epoch and band has a flux value and an associated uncertainty. 

\subsubsection{Photometric definition of RETs \label{subsubsec:trise_gr}}
To make an improved selection of RETs we use a parameter space described by observed $g-r$ colour and the time taken to rise from non-detection to peak $r$-band magnitude in the observer frame. Using rise times and colours from GP lightcurves, we populate this parameter space with the sample of 72 RETs from the \citetalias{Pursiainen2018} method, updated with 20 extra objects found using that method in the fifth season of DES. We add spectroscopically confirmed SNe of types Ia, Ibc, II, and SLSNe observed by DES in order to verify that they are rejected by the search method. We keep objects passing the following selection requirements (cuts):
\begin{enumerate}
\item The transient was detected in only one DES-SN observing season.
\item Maximum observed brightness in both $g$- and $r$-bands was brighter than 24 mag (in the eight `shallow' DES-SN fields) or 25 mag (in the two `deep' fields), as in \citetalias{Pursiainen2018}.
\end{enumerate}

 SNe Ia and RETs populate two distinct regions of $g-r$ vs. $t_{\mathrm{rise; r}}$ parameter space (Fig.~\ref{fig:selection}), where $t_{\mathrm{rise; r}}$ is the time to rise from non-detection to peak $r$-band brightness, and the $g-r$ colour is measured at peak brightness. RETs appear bluer and faster than the typical SNe. We define a region in this parameter space which minimises the contamination of non-RETs (purity) while maximising the total fraction of RETs (completeness). The resulting limits are $-1.5 < g-r < 0.6 $ and $t_{\mathrm{rise; r}} <16.5$, corresponding to the red box in Fig.~\ref{fig:selection}.

\subsubsection{Removal of active galactic nuclei \label{subsubsec:cnn}}

In order to apply selection criterion i) from Section \ref{subsubsec:trise_gr} it is necessary to distinguish between DES-SN candidates that are truly multi-season events (typically active galactic nuclei; AGN) and those that are single-season events with spurious detections in other seasons. 
To detect and filter out AGN we use a basic convolutional neural network (CNN) classifier. We train the CNN on spectroscopically confirmed SNe of all types as well as the 92 RETs identified using the \citetalias{Pursiainen2018} method (181 objects) and on spectroscopically typed AGN (182 objects), and use it to separate the sample into two photometric subtypes: AGN-like and SN-like. The classifier returns SNe-like objects with an accuracy of 0.992 on the test set (391 SNe and 79 AGN). No SNe were classified as AGN-like. The remaining AGN classified as SNe-like are removed by manual vetting later in the process. The CNN does not separate the SN-like objects into RETs and other SN subtypes: this is done in the subsequent processing step.

\subsubsection{Final DES RET sample}
The GP lightcurves of all DES-SN candidates are classified as AGN-like or SNe-like by the CNN, and are then subject to the lightcurve quality cuts i) and ii) (Section \ref{subsubsec:trise_gr}), resulting in 2259 objects, of which 939 lie inside the colour and rise-time region which we defined for RETs. These objects are subject to a further set of cuts in order to remove remaining contaminants. We impose a cut based on a SN classifier (\texttt{PSNID}; \citealt{Sako2008}), that returns a normalised goodness of fit to different SN Ia and CCSN templates, along with a Bayesian probability of it being a SN Ia. To remove highly-probable SNe Ia, we use threshold probabilities of P(Ia)$<0.91$ and P(Ia; Bayes)$<0.82$ respectively to the above algorithms, which removes 46 objects from the RET parameter space. In order to further remove longer-lived SNe, the decline time to half of the peak brightness must be $<24$ days. This removes 347 SNe, resulting in 546 objects remaining inside the parameter space. The final 546 transients have been visually inspected, with the majority rejected for clearly being spurious detections, obvious multi-season variability that was not picked up by the CNN, or showing a longer timescale decline. 

Using the above method recovers 87 of the 92 RETs found using the \citetalias{Pursiainen2018} technique, and adds a further 14. The five were not recovered as their GP lightcurves were fainter than the limits given above in either g or r band.
We refer to the resulting sample as DES RETs. Of the 106 objects in the sample, 96 have a host galaxy detected in deep host galaxy photometry of \citetalias{Wiseman2020} when using the Directional Light Radius method \citep{Sullivan2006} to associate hosts as per \citetalias{Wiseman2020}. Of these, 49 have a host galaxy spectroscopic redshift which we access through an internal release of the OzDES Global Redshift Catalog (GRC; v.2020\_01\_04). The full OzDES redshift catalogue will be available alongside the public data release detailed in \citet{Lidman2020}. A further three have redshifts obtained from narrow lines observed in spectra of the transients themselves. We do not consider these three objects for the analysis, since we are unable to separate transient and host contributions to the spectra.
A selection of the host galaxies is shown in Fig.~\ref{fig:mosaic}, centred on the location of the transient. The figure showcases the diversity of host galaxy morphologies and colours, while also displaying the limitations of ground-based observations of high-redshift, relatively small galaxies in terms of spatial resolution.
Fig.~\ref{fig:z_dist} shows the distribution of redshifts amongst the 49 hosts for which such a measurement was possible. The effect that the redshift selection function has on the results is discussed in Section \ref{subsec:disc_bias}.
The observational properties of the 96 detected hosts are displayed in Table \ref{tab:obs}. We highlight objects that were not presented in \citetalias{Pursiainen2018}. We also highlight a small subset of objects for which the redshift has taken on a new value than that presented in \citetalias{Pursiainen2018} due to further OzDES observations leading to a more accurate determination. 

\subsection{Comparison samples \label{subsec:comparison}}

In order to compare the host galaxies of DES RETs to those discovered in other surveys as well as other types of explosive transient, we draw upon samples in the literature. 

\subsubsection{RETs \label{subsubsec:compare_rets}}
Since the DES sample of RETs is by far the largest discovered to date, there is no other large sample of RETs with which to compare host galaxy properties. \citet{Drout2014} present host galaxies of 10 RETs discovered in the Pan-STARRS survey, with measurements of stellar masses, SFRs, and metallicities. We also compare with the host galaxy of SNLS04D4ec \citep{Arcavi2016}. To this we add the low-redshift transient AT2018cow (nicknamed ``The Cow''). The host galaxy of AT2018cow has been studied with photometric measurements \citep{Perley2019} as well as with an integral field spectrograph \citep{Lyman2020}, with consistent results. For our comparison, we use the galaxy-averaged stellar mass and SFR from \citet{Lyman2020} and the metallicity from the host nucleus as reported by both \citet{Morokuma-Matsui2019} and \citet{Lyman2020} as it best represents the method of obtaining spectra for our RET sample. We further compare to SN2018gep with data from  \citet{Ho2019}, and ZTF18abvkwla (nicknamed ``The Koala'') from \citet{Ho2020}.

\subsubsection{SNe and GRBs \label{subsubsec:compare_CCSNe}}

In compiling a set of comparison samples, we aim for the least biased selections possible. This requires surveys to be untargetted (they were not monitoring certain galaxies in order to search for SNe), ideally complete, and also covering a similar redshift range to the RETs. While in practice the second and third of these criteria are difficult to achieve, particularly with the fainter CCSNe, we are able to choose comparison samples from untargetted surveys to mitigate initial selection biases.

To compare with CCSNe, we draw on the untargetted sample of 47 SNe II from the Palomar Transient Factory (PTF; \citealt{Stoll2013}), which is likely complete in terms of hosts (all SNe have an associated host). While this sample lies at much lower redshift than the DES RETs (a maximum of 0.18 and mean of 0.05), redshift evolution is easier to account for in a less biased way than correcting for unknown incompleteness. We add to this the compilation of 56 untargeted SESNe from \citet{Sanders2012}, with a maximum redshift of 0.26 and a mean of 0.05. Since \citet{Sanders2012} do not report host galaxy magnitudes, stellar masses or SFRs, we cross-match the SN positions with the the Sloan Digital Sky Survey (SDSS; \citealt{York2000}) Data Release 16 (DR16; \citealt{Ahumada2019}) and perform our own SED fit using the method outlined in Section \ref{subsec:sedfit}. We are able to do this SED fit for 38 objects, with the others lying outside of the SDSS footprint.

We use the sample of GRB host galaxies of \citet{Kruehler2015}, using only galaxies with $z<1$ in order to maintain completeness, resulting in a sample of 29 hosts with a mean redshift of 0.66. To investigate similarities with SLSNe, we use the host galaxy sample from PTF presented in \citet{Perley2016c} with a mean redshift of 0.24 and a maximum of 0.50.

The host galaxy properties of the above samples are not all derived using the same methods. In terms of SED fitting, the largest systematic offsets in derived properties are due to differences in the assumed initial mass function (IMF). \citet{Stoll2013} and \citet{Drout2014} assume a \citet{Salpeter1955} IMF whereas all other samples considered (including those calculated in this work in Section \ref{subsec:sedfit}) are determined assuming a \citet{Chabrier2003} IMF. Stellar masses and star-formation rates derived using a Salpeter IMF are roughly $\sim 1.72$ times higher than those using a Chabrier IMF \citep{Speagle2014}, and we convert the Salpeter-derived values by this factor in order to compare them.

\subsection{Field Galaxies \label{subsubsec:sdss}}

To show how RETs compare to the galaxy population as a whole, we use a sample of $\sim800,000$ measurements from the MPA-JHU catalogues of stellar masses (based on the methods of \citealt{Kauffmann2003,Salim2007}), SFRs (based off \citealt{Brinchmann2004}), and metallicities (based off \citealt{Tremonti2004}) from the  catalogues of SDSS Data Relase 7 (DR7; \citealt{Abazajian2009}). The mean redshift is 0.08 which is much lower than for the RETs, such that significant evolution in the galaxy population has happened between the majority of RET hosts and the SDSS sample. We do not correct for this, but take it into account when analysing our findings. 
\section{Host galaxy observations}
\label{sec:obs}
\subsection{Photometry \label{subsec:phot}}

The host galaxy photometry for the sample of RETs is taken from the catalogue of \citetalias{Wiseman2020}, which is based upon deep coadds reaching $r$-band limiting magnitudes of 26.5. The coadds were created using data from all five seasons of DES-SN, but by excluding one season at a time in order for that coadd not to include contamination from the transients in that season. For this sample, the limiting magnitude for obtaining a spectroscopic redshift (Section \ref{subsec:spec}) is $\sim 24.5$, meaning that all hosts in the sample are detected with a high S/N.

\subsection{Spectroscopy \label{subsec:spec}}
Accurate redshifts for DES-SN were obtained by OzDES, a dedicated DES spectroscopic follow-up campaign based at the $3.9~\textrm{m}$ Anglo-Australian Telescope (AAT) using the AAOmega fibre-fed spectrograph and 2dF fibre positioner. The observation strategy of OzDES was to point at one of the ten DES-SN fields, and place fibres at the positions of transient hosts, continually coadding the spectra of a particular host until a redshift was obtained at which point the fibre could be allocated to a different transient. The spectra have a resolution of 1400-1700 and a wavelength range of $3700 -8800$~\AA, and are reduced using the OzDES pipeline which makes use of a modified version of v6.46 of the 2dfdr \citep{Croom2004} along with internal scripts. We use internal data release 7, a preliminary version of the public data release which is detailed in \citet{Lidman2020}. Extensive description and discussion of OzDES can be found in \citet{Yuan2015,Childress2017,Lidman2020}. We also obtained redshifts for some transient hosts serendipitously as part of the Looking at the Distant Universe with the MeerKAT Array (LADUMA) survey\footnote{\url{http://www.laduma.uct.ac.za}}, three of which are present in our sample.
Objects for which the host already had a publicly available redshift were not observed with OzDES, but merged into the GRC nontheless. Surveys fulfilling this criteria include the Australia Telescope Large Area Survey (ATLAS; 
citealt{Mao2012}, the Arizona CDFS Environment Survey (ACES; \citealt{Cooper2012}), and the PRIsm MUlti-object Survey (PRIMUS; \citealt{Coil2011,Cool2013}). Where the spectra from which those redshifts were derived are also public they are included in this analysis. These comprise the Galaxy and Mass Assembly survey (GAMA \citealt{Driver2009,Baldry2018}) and SDSS. In total we analyse 45 spectra, with a mean continuum signal-to-noise ratio (SNR) of 2.56/pixel. We stress that the emission lines are detected with a higher SNR than this.

\begin{figure*}
\includegraphics[width=\textwidth]{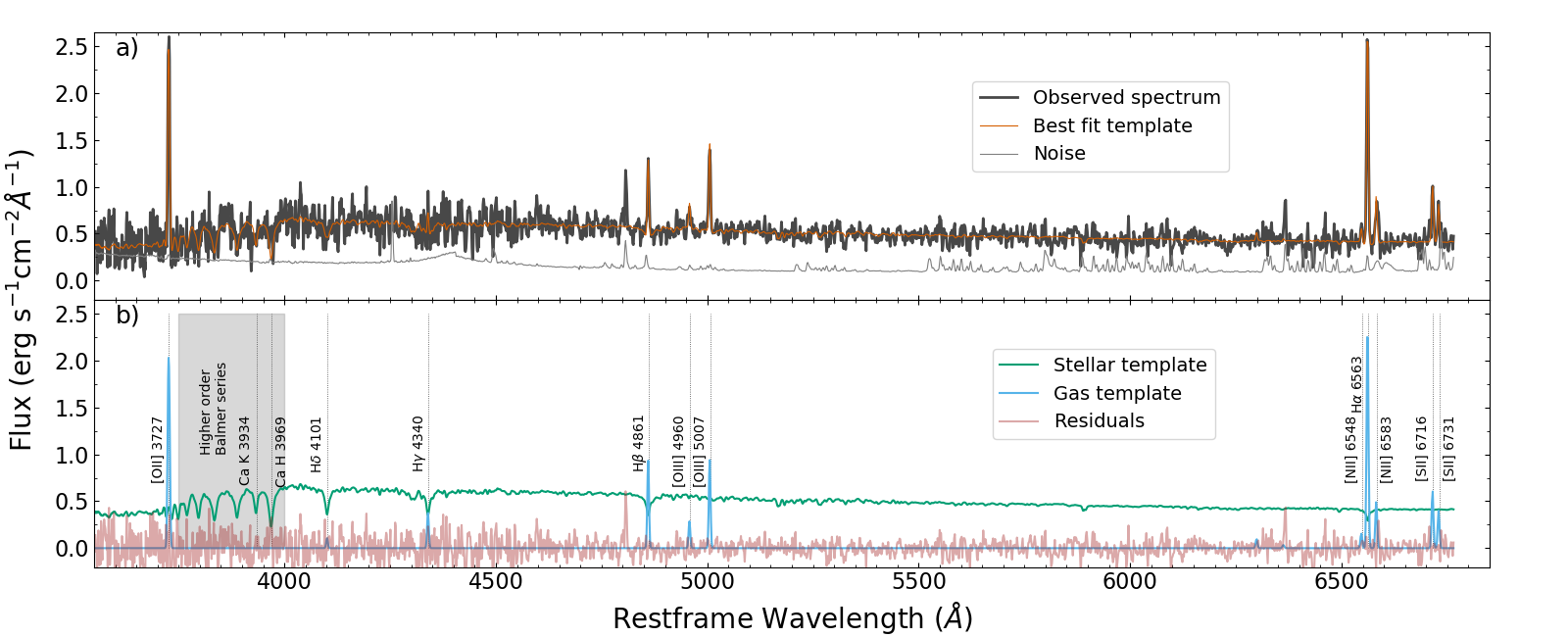}
\caption{The spectrum of DES16C2ggt, decomposed into its constituent components according to the \texttt{pPXF} fit. a) the best fit superimposed on the observed spectrum. b) the constituent parts of the decomposition: the stellar template including absorption features, and the nebular gas emission. The grey area shows the location of higher-order (\hepsilon onwards) Balmer lines.
\label{fig:host_spec}}
\end{figure*}

\begin{figure}
\includegraphics[width=0.5\textwidth]{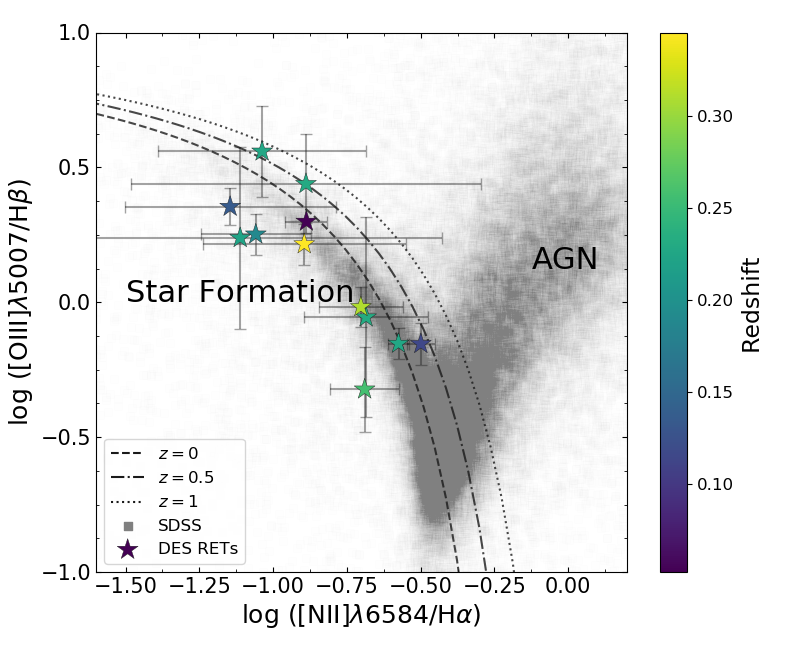}
\caption{Baldwin-Phillips-Terlovich (BPT) diagram for RET hosts, showing that the emission lines are consistent with being generated by star formation rather than AGN activity. The three curves show the delimitation between star formation and AGN at $z =0, 0.5, 1$ according to \citet{Kewley2013}. The redshift of each RET is indicated by the colourbar on the right.
\label{fig:bpt}}
\end{figure}

\section{Estimating host galaxy properties}
\label{sec:measure}
\subsection{Photometric stellar parameters \label{subsec:sedfit}}
To estimate the physical properties of the DES RET host galaxies, we generate synthetic photometry in the DES $griz$ bands by combining the individual SEDs of simple stellar population models. We simulate a suite of synthetic galaxy star-formation histories from which we synthesise model SEDs using stellar population models from \citet{Bruzual2003}  and a \citet{Chabrier2003} initial mass function (IMF).  The suite of models is drawn from the same distribution of parameters as used in \citet{Kauffmann2003} and similar papers \citep[e.g.][]{Gallazzi2005,Gallazzi2009} and closely follows the method of \citet{Childress2013}. From the synthetic SEDs we derive model magnitudes in the DES $griz$ bands and compare them to the observed Kron magnitudes from the deep photometric catalogue of \citetalias{Wiseman2020}. For each set of model and observed magnitudes we calculate a $\chi^2$ value, and from these estimate a probability density function (PDF) for key model parameters (mass-to-light ratio $M/L$, from which we derive $M_*$, and specific star-formation rate sSFR). To estimate uncertainties, we take the values at the 16th and 84th percentiles of the resulting PDF to be our $1\sigma$ lower and upper bounds. The results are presented in Table \ref{tab:derived}. These parameters are estimated using global photometry of the entire galaxies, and thus represent the overall stellar population.

\subsection{Spectroscopic gas-phase parameters \label{subsec:linefit}}

To estimate parameters from the OzDES host galaxy spectra requires several processing steps. We first apply a flux calibration by `mangling' the spectrum such that the integrated flux over the wavelength ranges of the DES photometric bands matches that measured in the photometry (further details on the mangling process are provided in \citealt{Swann2020}). We use a circular aperture of diameter 2\arcsec, matching the size of the spectrograph fibres. The resulting spectrum is a more accurate representation of the true spectrum at that point in the galaxy. This only holds, however, for the area covered by the fibre and we note that the resulting spectrum is not necessarily representative of the galaxy as a whole. There are several reasons this measurement may differ from one made at the SN explosion site, such as metallicity and age gradients or structure such as bars and discs (see e.g. \citealt{Iglesias-Paramo2013,Iglesias-Paramo2016}). We proceed with our analysis with this caveat acknowledged.

In order to subtract the stellar component of the host galaxy spectra, we use the Penalized PiXel-Fitting software (\texttt{pPXF}; \citealt{Cappellari2004,Cappellari2012,Cappellari2017}), using the MILES library of single stellar populations \citep{Vazdekis2010}. By subtracting the best-fitting composite stellar spectrum from the \texttt{pPXF} fit, we are left with a `gas' spectrum, comprising the emission lines. An example of this procedure is shown in Fig.~\ref{fig:host_spec}. We fit the emission lines with Gaussian profiles. In order to estimate the uncertainty on the emission line fluxes, we fit $10^4$ realisations of the line, each time adding perturbations to the line by drawing from a Gaussian distribution based on the variance spectrum. We take the mean and standard deviation of the resulting fits as our flux and its uncertainty, respectively. Line fluxes are presented in Table \ref{tab:fluxes}.

\begin{table*}

\caption{Host galaxy properties for the 49 DES RET host galaxies with redshifts and host galaxy spectra. The table is available in the online version in a machine readable format.}
\begin{threeparttable}
\begin{tabular}{lrrrllllll}
\toprule
Transient Name &$\log \left(M_*\right)$ &$\log \left(\mathrm{SFR}\right)$ &  $\log \left(\mathrm{sSFR}\right)$ &                                                           \multicolumn{6}{c}{$12+\log\left(\mathrm{O/H}\right) $}\\
{} & $\left( M_{\odot}\right)$ & $\left(M_{\odot} \mathrm{yr}^{-1}\right)$&$\left(\mathrm{yr}^{-1} \right)$& Best\tnote{a} &                                                           D16 &                                                       PP04 N2 &                                                     PP04 O3N2 &                                                      KK04 R23 &                                                   Average O3N2\tnote{b}\\
\midrule
DES13C1tgd  &  $10.24 _{\scriptscriptstyle 0.09} ^{\scriptscriptstyle 0.08}$ &   $0.10 _{\scriptscriptstyle 0.47} ^{\scriptscriptstyle 0.38}$ &  $-10.14 _{\scriptscriptstyle 0.39} ^{\scriptscriptstyle 0.30}$ &  $8.67 _{\scriptscriptstyle 0.19} ^{\scriptscriptstyle 0.18}$ &  $8.62 _{\scriptscriptstyle 0.20} ^{\scriptscriptstyle 0.18}$ &  $8.72 _{\scriptscriptstyle 0.16} ^{\scriptscriptstyle 0.16}$ &                                                             - &                                                             - &     $8.80 _{\scriptscriptstyle 0.18} ^{\scriptscriptstyle -}$ \\
DES13C3bcok &  $11.57 _{\scriptscriptstyle 0.11} ^{\scriptscriptstyle 0.11}$ &   $1.20 _{\scriptscriptstyle 0.50} ^{\scriptscriptstyle 0.49}$ &  $-10.37 _{\scriptscriptstyle 0.39} ^{\scriptscriptstyle 0.38}$ &  $8.99 _{\scriptscriptstyle 0.29} ^{\scriptscriptstyle 0.08}$ &                                                             - &                                                             - &                                                             - &  $8.99 _{\scriptscriptstyle 0.29} ^{\scriptscriptstyle 0.08}$ &  $8.66 _{\scriptscriptstyle 0.31} ^{\scriptscriptstyle 0.10}$ \\
DES13C3uig  &  $10.28 _{\scriptscriptstyle 0.10} ^{\scriptscriptstyle 0.23}$ &   $0.59 _{\scriptscriptstyle 0.29} ^{\scriptscriptstyle 0.40}$ &   $-9.68 _{\scriptscriptstyle 0.19} ^{\scriptscriptstyle 0.16}$ &                                                             - &                                                             - &                                                             - &                                                             - &                                                             - &                                                             - \\
DES13E2lpk  &  $10.66 _{\scriptscriptstyle 0.09} ^{\scriptscriptstyle 0.10}$ &   $0.46 _{\scriptscriptstyle 0.36} ^{\scriptscriptstyle 0.45}$ &  $-10.20 _{\scriptscriptstyle 0.27} ^{\scriptscriptstyle 0.35}$ &  $8.89 _{\scriptscriptstyle 0.14} ^{\scriptscriptstyle 0.09}$ &                                                             - &                                                             - &                                                             - &  $8.89 _{\scriptscriptstyle 0.14} ^{\scriptscriptstyle 0.09}$ &  $8.54 _{\scriptscriptstyle 0.15} ^{\scriptscriptstyle 0.11}$ \\
DES13S2wxf  &   $9.86 _{\scriptscriptstyle 0.03} ^{\scriptscriptstyle 0.06}$ &   $0.29 _{\scriptscriptstyle 0.07} ^{\scriptscriptstyle 0.10}$ &   $-9.57 _{\scriptscriptstyle 0.04} ^{\scriptscriptstyle 0.04}$ &  $7.62 _{\scriptscriptstyle 0.70} ^{\scriptscriptstyle 0.62}$ &                                                             - &                                                             - &                                                             - &  $7.62 _{\scriptscriptstyle 0.70} ^{\scriptscriptstyle 0.62}$ &                                                             - \\
DES13X1hav  &   $9.16 _{\scriptscriptstyle 0.16} ^{\scriptscriptstyle 0.38}$ &  $-0.44 _{\scriptscriptstyle 0.37} ^{\scriptscriptstyle 0.93}$ &   $-9.60 _{\scriptscriptstyle 0.21} ^{\scriptscriptstyle 0.55}$ &  $8.62 _{\scriptscriptstyle 0.46} ^{\scriptscriptstyle 0.24}$ &                                                             - &                                                             - &                                                             - &  $8.62 _{\scriptscriptstyle 0.46} ^{\scriptscriptstyle 0.24}$ &  $8.28 _{\scriptscriptstyle 0.24} ^{\scriptscriptstyle 0.24}$ \\
DES13X2wvv  &   $9.79 _{\scriptscriptstyle 0.14} ^{\scriptscriptstyle 0.34}$ &   $0.19 _{\scriptscriptstyle 0.43} ^{\scriptscriptstyle 1.25}$ &   $-9.60 _{\scriptscriptstyle 0.28} ^{\scriptscriptstyle 0.91}$ &  $8.68 _{\scriptscriptstyle 0.27} ^{\scriptscriptstyle 0.15}$ &                                                             - &                                                             - &                                                             - &  $8.68 _{\scriptscriptstyle 0.27} ^{\scriptscriptstyle 0.15}$ &  $8.33 _{\scriptscriptstyle 0.17} ^{\scriptscriptstyle 0.16}$ \\
DES13X3gmd  &  $10.33 _{\scriptscriptstyle 0.50} ^{\scriptscriptstyle 0.44}$ &   $0.74 _{\scriptscriptstyle 0.75} ^{\scriptscriptstyle 0.86}$ &   $-9.59 _{\scriptscriptstyle 0.25} ^{\scriptscriptstyle 0.42}$ &                                                             - &                                                             - &                                                             - &                                                             - &                                                             - &                                                             - \\
DES13X3gms  &   $9.41 _{\scriptscriptstyle 0.16} ^{\scriptscriptstyle 0.27}$ &  $-0.06 _{\scriptscriptstyle 0.51} ^{\scriptscriptstyle 0.78}$ &   $-9.47 _{\scriptscriptstyle 0.35} ^{\scriptscriptstyle 0.52}$ &  $8.51 _{\scriptscriptstyle 0.96} ^{\scriptscriptstyle 0.50}$ &                                                             - &                                                             - &                                                             - &  $8.51 _{\scriptscriptstyle 0.96} ^{\scriptscriptstyle 0.50}$ &  $8.21 _{\scriptscriptstyle 0.48} ^{\scriptscriptstyle 0.48}$ \\
DES13X3npb  &  $11.02 _{\scriptscriptstyle 0.31} ^{\scriptscriptstyle 0.18}$ &   $1.14 _{\scriptscriptstyle 0.70} ^{\scriptscriptstyle 1.35}$ &   $-9.88 _{\scriptscriptstyle 0.39} ^{\scriptscriptstyle 1.17}$ &  $8.72 _{\scriptscriptstyle 0.37} ^{\scriptscriptstyle 0.23}$ &                                                             - &                                                             - &                                                             - &  $8.72 _{\scriptscriptstyle 0.37} ^{\scriptscriptstyle 0.23}$ &  $8.37 _{\scriptscriptstyle 0.22} ^{\scriptscriptstyle 0.26}$ \\
DES13X3nyg  &   $9.31 _{\scriptscriptstyle 0.26} ^{\scriptscriptstyle 0.42}$ &   $0.05 _{\scriptscriptstyle 0.53} ^{\scriptscriptstyle 0.73}$ &   $-9.27 _{\scriptscriptstyle 0.27} ^{\scriptscriptstyle 0.32}$ &  $7.69 _{\scriptscriptstyle 0.72} ^{\scriptscriptstyle 0.61}$ &                                                             - &                                                             - &                                                             - &  $7.69 _{\scriptscriptstyle 0.72} ^{\scriptscriptstyle 0.61}$ &                                                             - \\
DES14C3tnz  &  $10.21 _{\scriptscriptstyle 0.22} ^{\scriptscriptstyle 0.41}$ &   $0.78 _{\scriptscriptstyle 0.60} ^{\scriptscriptstyle 0.88}$ &   $-9.44 _{\scriptscriptstyle 0.38} ^{\scriptscriptstyle 0.47}$ &  $8.38 _{\scriptscriptstyle 0.85} ^{\scriptscriptstyle 0.42}$ &                                                             - &                                                             - &                                                             - &  $8.38 _{\scriptscriptstyle 0.85} ^{\scriptscriptstyle 0.42}$ &  $8.15 _{\scriptscriptstyle 0.29} ^{\scriptscriptstyle 0.29}$ \\
DES14C3tvw  &  $11.09 _{\scriptscriptstyle 0.11} ^{\scriptscriptstyle 0.09}$ &   $1.02 _{\scriptscriptstyle 0.46} ^{\scriptscriptstyle 0.39}$ &  $-10.07 _{\scriptscriptstyle 0.35} ^{\scriptscriptstyle 0.29}$ &                                                             - &                                                             - &                                                             - &                                                             - &                                                             - &                                                             - \\
DES14S2anq  &   $9.41 _{\scriptscriptstyle 0.10} ^{\scriptscriptstyle 0.23}$ &  $-0.34 _{\scriptscriptstyle 0.43} ^{\scriptscriptstyle 0.54}$ &   $-9.75 _{\scriptscriptstyle 0.33} ^{\scriptscriptstyle 0.32}$ &  $8.34 _{\scriptscriptstyle 0.15} ^{\scriptscriptstyle 0.02}$ &  $8.16 _{\scriptscriptstyle 0.11} ^{\scriptscriptstyle 0.09}$ &  $8.34 _{\scriptscriptstyle 0.04} ^{\scriptscriptstyle 0.04}$ &  $8.35 _{\scriptscriptstyle 0.02} ^{\scriptscriptstyle 0.02}$ &  $8.36 _{\scriptscriptstyle 0.01} ^{\scriptscriptstyle 0.01}$ &  $8.29 _{\scriptscriptstyle 0.02} ^{\scriptscriptstyle 0.02}$ \\
DES14S2plb  &  $10.36 _{\scriptscriptstyle 0.23} ^{\scriptscriptstyle 0.25}$ &   $0.89 _{\scriptscriptstyle 0.64} ^{\scriptscriptstyle 0.77}$ &   $-9.47 _{\scriptscriptstyle 0.40} ^{\scriptscriptstyle 0.52}$ &  $8.65 _{\scriptscriptstyle 0.05} ^{\scriptscriptstyle 0.29}$ &  $8.65 _{\scriptscriptstyle 0.07} ^{\scriptscriptstyle 0.06}$ &  $8.63 _{\scriptscriptstyle 0.05} ^{\scriptscriptstyle 0.05}$ &  $8.62 _{\scriptscriptstyle 0.03} ^{\scriptscriptstyle 0.03}$ &  $8.95 _{\scriptscriptstyle 0.05} ^{\scriptscriptstyle 0.04}$ &  $8.64 _{\scriptscriptstyle 0.05} ^{\scriptscriptstyle 0.04}$ \\
DES14S2pli  &  $10.27 _{\scriptscriptstyle 0.11} ^{\scriptscriptstyle 0.34}$ &   $0.50 _{\scriptscriptstyle 0.38} ^{\scriptscriptstyle 0.61}$ &   $-9.77 _{\scriptscriptstyle 0.27} ^{\scriptscriptstyle 0.27}$ &  $8.95 _{\scriptscriptstyle 0.08} ^{\scriptscriptstyle 0.06}$ &                                                             - &                                                             - &                                                             - &  $8.95 _{\scriptscriptstyle 0.08} ^{\scriptscriptstyle 0.06}$ &  $8.61 _{\scriptscriptstyle 0.09} ^{\scriptscriptstyle 0.07}$ \\
DES14X1bnh  &  $11.71 _{\scriptscriptstyle 0.69} ^{\scriptscriptstyle 0.25}$ &   $1.53 _{\scriptscriptstyle 0.95} ^{\scriptscriptstyle 0.52}$ &  $-10.18 _{\scriptscriptstyle 0.25} ^{\scriptscriptstyle 0.26}$ &                                                             - &                                                             - &                                                             - &                                                             - &                                                             - &                                                             - \\
DES14X3pkl  &   $9.60 _{\scriptscriptstyle 0.34} ^{\scriptscriptstyle 0.26}$ &   $0.26 _{\scriptscriptstyle 0.85} ^{\scriptscriptstyle 0.72}$ &   $-9.34 _{\scriptscriptstyle 0.51} ^{\scriptscriptstyle 0.46}$ &  $8.38 _{\scriptscriptstyle 0.29} ^{\scriptscriptstyle 0.20}$ &  $8.22 _{\scriptscriptstyle 0.30} ^{\scriptscriptstyle 0.23}$ &  $8.48 _{\scriptscriptstyle 0.16} ^{\scriptscriptstyle 0.15}$ &                                                             - &                                                             - &  $8.52 _{\scriptscriptstyle 0.18} ^{\scriptscriptstyle 0.18}$ \\
DES15C2eal  &   $8.42 _{\scriptscriptstyle 0.09} ^{\scriptscriptstyle 0.18}$ &  $-1.29 _{\scriptscriptstyle 0.30} ^{\scriptscriptstyle 0.33}$ &   $-9.72 _{\scriptscriptstyle 0.21} ^{\scriptscriptstyle 0.15}$ &  $8.30 _{\scriptscriptstyle 0.31} ^{\scriptscriptstyle 0.25}$ &  $8.04 _{\scriptscriptstyle 0.51} ^{\scriptscriptstyle 0.32}$ &  $8.32 _{\scriptscriptstyle 0.19} ^{\scriptscriptstyle 0.19}$ &  $8.34 _{\scriptscriptstyle 0.20} ^{\scriptscriptstyle 0.15}$ &  $8.39 _{\scriptscriptstyle 0.34} ^{\scriptscriptstyle 0.50}$ &  $8.28 _{\scriptscriptstyle 0.16} ^{\scriptscriptstyle 0.25}$ \\
DES15C3lpq  &   $9.32 _{\scriptscriptstyle 0.20} ^{\scriptscriptstyle 0.39}$ &   $0.86 _{\scriptscriptstyle 0.73} ^{\scriptscriptstyle 1.08}$ &   $-8.45 _{\scriptscriptstyle 0.53} ^{\scriptscriptstyle 0.68}$ &  $8.30 _{\scriptscriptstyle 0.58} ^{\scriptscriptstyle 0.33}$ &                                                             - &                                                             - &                                                             - &  $8.30 _{\scriptscriptstyle 0.58} ^{\scriptscriptstyle 0.33}$ &  $8.15 _{\scriptscriptstyle 0.14} ^{\scriptscriptstyle 0.14}$ \\
DES15C3lzm  &  $10.06 _{\scriptscriptstyle 0.20} ^{\scriptscriptstyle 0.28}$ &   $1.08 _{\scriptscriptstyle 0.68} ^{\scriptscriptstyle 0.75}$ &   $-8.98 _{\scriptscriptstyle 0.48} ^{\scriptscriptstyle 0.47}$ &                                                             - &                                                             - &                                                             - &                                                             - &                                                             - &                                                             - \\
DES15C3mem  &  $10.76 _{\scriptscriptstyle 0.07} ^{\scriptscriptstyle 0.37}$ &   $1.04 _{\scriptscriptstyle 0.37} ^{\scriptscriptstyle 0.64}$ &   $-9.72 _{\scriptscriptstyle 0.31} ^{\scriptscriptstyle 0.27}$ &                                                             - &                                                             - &                                                             - &                                                             - &                                                             - &                                                             - \\
DES15C3mgq  &   $8.41 _{\scriptscriptstyle 0.12} ^{\scriptscriptstyle 0.29}$ &  $-1.12 _{\scriptscriptstyle 0.37} ^{\scriptscriptstyle 0.55}$ &   $-9.53 _{\scriptscriptstyle 0.25} ^{\scriptscriptstyle 0.26}$ &  $8.37 _{\scriptscriptstyle 0.27} ^{\scriptscriptstyle 0.29}$ &  $8.45 _{\scriptscriptstyle 0.50} ^{\scriptscriptstyle 0.48}$ &  $8.43 _{\scriptscriptstyle 0.23} ^{\scriptscriptstyle 0.24}$ &  $8.34 _{\scriptscriptstyle 0.17} ^{\scriptscriptstyle 0.11}$ &  $8.34 _{\scriptscriptstyle 0.34} ^{\scriptscriptstyle 0.39}$ &  $8.32 _{\scriptscriptstyle 0.14} ^{\scriptscriptstyle 0.20}$ \\
DES15C3nat  &  $10.49 _{\scriptscriptstyle 0.17} ^{\scriptscriptstyle 0.58}$ &   $1.14 _{\scriptscriptstyle 0.60} ^{\scriptscriptstyle 0.98}$ &   $-9.34 _{\scriptscriptstyle 0.43} ^{\scriptscriptstyle 0.41}$ &                                                             - &                                                             - &                                                             - &                                                             - &                                                             - &                                                             - \\
DES15C3opk  &   $9.89 _{\scriptscriptstyle 0.44} ^{\scriptscriptstyle 0.22}$ &   $0.89 _{\scriptscriptstyle 1.20} ^{\scriptscriptstyle 0.67}$ &   $-9.01 _{\scriptscriptstyle 0.75} ^{\scriptscriptstyle 0.45}$ &  $8.57 _{\scriptscriptstyle 0.80} ^{\scriptscriptstyle 0.33}$ &                                                             - &                                                             - &                                                             - &  $8.57 _{\scriptscriptstyle 0.80} ^{\scriptscriptstyle 0.33}$ &  $8.24 _{\scriptscriptstyle 0.31} ^{\scriptscriptstyle 0.31}$ \\
DES15C3opp  &   $9.15 _{\scriptscriptstyle 0.21} ^{\scriptscriptstyle 0.32}$ &   $0.13 _{\scriptscriptstyle 0.78} ^{\scriptscriptstyle 0.84}$ &   $-9.02 _{\scriptscriptstyle 0.57} ^{\scriptscriptstyle 0.52}$ &  $8.45 _{\scriptscriptstyle 0.59} ^{\scriptscriptstyle 0.32}$ &                                                             - &                                                             - &                                                             - &  $8.45 _{\scriptscriptstyle 0.59} ^{\scriptscriptstyle 0.32}$ &  $8.17 _{\scriptscriptstyle 0.24} ^{\scriptscriptstyle 0.24}$ \\
DES15E2nqh  &   $9.28 _{\scriptscriptstyle 0.27} ^{\scriptscriptstyle 0.28}$ &   $0.11 _{\scriptscriptstyle 0.72} ^{\scriptscriptstyle 0.94}$ &   $-9.17 _{\scriptscriptstyle 0.45} ^{\scriptscriptstyle 0.66}$ &  $8.83 _{\scriptscriptstyle 0.52} ^{\scriptscriptstyle 0.15}$ &                                                             - &                                                             - &                                                             - &  $8.83 _{\scriptscriptstyle 0.52} ^{\scriptscriptstyle 0.15}$ &  $8.48 _{\scriptscriptstyle 0.34} ^{\scriptscriptstyle 0.17}$ \\
DES15S1fli  &  $10.24 _{\scriptscriptstyle 0.27} ^{\scriptscriptstyle 0.26}$ &   $1.54 _{\scriptscriptstyle 1.01} ^{\scriptscriptstyle 0.67}$ &   $-8.70 _{\scriptscriptstyle 0.75} ^{\scriptscriptstyle 0.41}$ &  $8.86 _{\scriptscriptstyle 0.08} ^{\scriptscriptstyle 0.06}$ &                                                             - &                                                             - &                                                             - &  $8.86 _{\scriptscriptstyle 0.08} ^{\scriptscriptstyle 0.06}$ &  $8.52 _{\scriptscriptstyle 0.08} ^{\scriptscriptstyle 0.07}$ \\
DES15S1fll  &   $9.20 _{\scriptscriptstyle 0.20} ^{\scriptscriptstyle 0.32}$ &  $-0.14 _{\scriptscriptstyle 0.64} ^{\scriptscriptstyle 0.54}$ &   $-9.34 _{\scriptscriptstyle 0.43} ^{\scriptscriptstyle 0.22}$ &  $8.26 _{\scriptscriptstyle 0.20} ^{\scriptscriptstyle 0.23}$ &  $8.37 _{\scriptscriptstyle 0.45} ^{\scriptscriptstyle 0.36}$ &  $8.28 _{\scriptscriptstyle 0.16} ^{\scriptscriptstyle 0.13}$ &  $8.22 _{\scriptscriptstyle 0.14} ^{\scriptscriptstyle 0.10}$ &  $8.26 _{\scriptscriptstyle 0.22} ^{\scriptscriptstyle 0.34}$ &  $8.22 _{\scriptscriptstyle 0.13} ^{\scriptscriptstyle 0.12}$ \\
DES15X2ead  &   $9.92 _{\scriptscriptstyle 0.08} ^{\scriptscriptstyle 0.20}$ &   $0.22 _{\scriptscriptstyle 0.41} ^{\scriptscriptstyle 0.47}$ &   $-9.69 _{\scriptscriptstyle 0.32} ^{\scriptscriptstyle 0.27}$ &  $8.45 _{\scriptscriptstyle 0.33} ^{\scriptscriptstyle 0.23}$ &  $8.22 _{\scriptscriptstyle 0.34} ^{\scriptscriptstyle 0.25}$ &  $8.47 _{\scriptscriptstyle 0.16} ^{\scriptscriptstyle 0.15}$ &  $8.52 _{\scriptscriptstyle 0.15} ^{\scriptscriptstyle 0.14}$ &  $8.61 _{\scriptscriptstyle 0.77} ^{\scriptscriptstyle 0.38}$ &  $8.43 _{\scriptscriptstyle 0.09} ^{\scriptscriptstyle 0.24}$ \\
DES15X3mxf  &   $9.93 _{\scriptscriptstyle 0.30} ^{\scriptscriptstyle 0.32}$ &   $0.78 _{\scriptscriptstyle 1.18} ^{\scriptscriptstyle 0.90}$ &   $-9.15 _{\scriptscriptstyle 0.88} ^{\scriptscriptstyle 0.58}$ &  $8.72 _{\scriptscriptstyle 0.25} ^{\scriptscriptstyle 0.15}$ &                                                             - &                                                             - &                                                             - &  $8.72 _{\scriptscriptstyle 0.25} ^{\scriptscriptstyle 0.15}$ &  $8.36 _{\scriptscriptstyle 0.18} ^{\scriptscriptstyle 0.16}$ \\
DES16C1cbd  &  $10.77 _{\scriptscriptstyle 0.15} ^{\scriptscriptstyle 0.22}$ &  $-1.43 _{\scriptscriptstyle 1.94} ^{\scriptscriptstyle 1.48}$ &  $-12.20 _{\scriptscriptstyle 1.79} ^{\scriptscriptstyle 1.26}$ &  $8.79 _{\scriptscriptstyle 0.37} ^{\scriptscriptstyle 0.20}$ &                                                             - &                                                             - &                                                             - &  $8.79 _{\scriptscriptstyle 0.37} ^{\scriptscriptstyle 0.20}$ &  $8.44 _{\scriptscriptstyle 0.28} ^{\scriptscriptstyle 0.22}$ \\
DES16C2ggt  &   $9.86 _{\scriptscriptstyle 0.29} ^{\scriptscriptstyle 0.36}$ &   $0.13 _{\scriptscriptstyle 0.46} ^{\scriptscriptstyle 0.94}$ &   $-9.73 _{\scriptscriptstyle 0.18} ^{\scriptscriptstyle 0.59}$ &  $8.50 _{\scriptscriptstyle 0.20} ^{\scriptscriptstyle 0.33}$ &  $8.28 _{\scriptscriptstyle 0.22} ^{\scriptscriptstyle 0.17}$ &  $8.45 _{\scriptscriptstyle 0.11} ^{\scriptscriptstyle 0.10}$ &  $8.51 _{\scriptscriptstyle 0.06} ^{\scriptscriptstyle 0.05}$ &  $8.85 _{\scriptscriptstyle 0.07} ^{\scriptscriptstyle 0.06}$ &  $8.50 _{\scriptscriptstyle 0.09} ^{\scriptscriptstyle 0.08}$ \\
DES16C3axz  &   $9.85 _{\scriptscriptstyle 0.19} ^{\scriptscriptstyle 0.33}$ &   $0.20 _{\scriptscriptstyle 0.56} ^{\scriptscriptstyle 0.78}$ &   $-9.65 _{\scriptscriptstyle 0.37} ^{\scriptscriptstyle 0.44}$ &  $8.59 _{\scriptscriptstyle 0.06} ^{\scriptscriptstyle 0.37}$ &  $8.53 _{\scriptscriptstyle 0.05} ^{\scriptscriptstyle 0.05}$ &  $8.56 _{\scriptscriptstyle 0.03} ^{\scriptscriptstyle 0.03}$ &  $8.59 _{\scriptscriptstyle 0.02} ^{\scriptscriptstyle 0.02}$ &  $8.97 _{\scriptscriptstyle 0.04} ^{\scriptscriptstyle 0.03}$ &  $8.61 _{\scriptscriptstyle 0.03} ^{\scriptscriptstyle 0.03}$ \\
DES16C3gin  &   $9.85 _{\scriptscriptstyle 0.27} ^{\scriptscriptstyle 0.32}$ &   $0.45 _{\scriptscriptstyle 0.68} ^{\scriptscriptstyle 1.00}$ &   $-9.41 _{\scriptscriptstyle 0.41} ^{\scriptscriptstyle 0.68}$ &  $8.84 _{\scriptscriptstyle 0.23} ^{\scriptscriptstyle 0.12}$ &                                                             - &                                                             - &                                                             - &  $8.84 _{\scriptscriptstyle 0.23} ^{\scriptscriptstyle 0.12}$ &  $8.49 _{\scriptscriptstyle 0.22} ^{\scriptscriptstyle 0.14}$ \\
DES16E2pv   &   $9.57 _{\scriptscriptstyle 0.12} ^{\scriptscriptstyle 0.43}$ &   $0.07 _{\scriptscriptstyle 0.39} ^{\scriptscriptstyle 0.98}$ &   $-9.50 _{\scriptscriptstyle 0.26} ^{\scriptscriptstyle 0.55}$ &  $8.93 _{\scriptscriptstyle 0.34} ^{\scriptscriptstyle 0.11}$ &                                                             - &                                                             - &                                                             - &  $8.93 _{\scriptscriptstyle 0.34} ^{\scriptscriptstyle 0.11}$ &  $8.59 _{\scriptscriptstyle 0.33} ^{\scriptscriptstyle 0.13}$ \\
DES16S1bbp  &   $9.06 _{\scriptscriptstyle 0.39} ^{\scriptscriptstyle 0.30}$ &   $0.30 _{\scriptscriptstyle 0.86} ^{\scriptscriptstyle 0.79}$ &   $-8.75 _{\scriptscriptstyle 0.48} ^{\scriptscriptstyle 0.49}$ &  $8.25 _{\scriptscriptstyle 0.17} ^{\scriptscriptstyle 0.10}$ &  $8.11 _{\scriptscriptstyle 0.33} ^{\scriptscriptstyle 0.25}$ &  $8.25 _{\scriptscriptstyle 0.09} ^{\scriptscriptstyle 0.07}$ &  $8.31 _{\scriptscriptstyle 0.08} ^{\scriptscriptstyle 0.05}$ &  $8.24 _{\scriptscriptstyle 0.12} ^{\scriptscriptstyle 0.13}$ &  $8.24 _{\scriptscriptstyle 0.06} ^{\scriptscriptstyle 0.04}$ \\
DES16S1dxu  &   $8.60 _{\scriptscriptstyle 0.28} ^{\scriptscriptstyle 0.41}$ &  $-0.49 _{\scriptscriptstyle 0.71} ^{\scriptscriptstyle 1.70}$ &   $-9.09 _{\scriptscriptstyle 0.43} ^{\scriptscriptstyle 1.29}$ &  $8.26 _{\scriptscriptstyle 0.15} ^{\scriptscriptstyle 0.12}$ &  $8.33 _{\scriptscriptstyle 0.45} ^{\scriptscriptstyle 0.38}$ &  $8.24 _{\scriptscriptstyle 0.14} ^{\scriptscriptstyle 0.11}$ &  $8.27 _{\scriptscriptstyle 0.13} ^{\scriptscriptstyle 0.07}$ &  $8.27 _{\scriptscriptstyle 0.10} ^{\scriptscriptstyle 0.10}$ &  $8.22 _{\scriptscriptstyle 0.12} ^{\scriptscriptstyle 0.07}$ \\
DES16X1eho  &  $10.70 _{\scriptscriptstyle 0.24} ^{\scriptscriptstyle 0.42}$ &   $0.01 _{\scriptscriptstyle 0.50} ^{\scriptscriptstyle 1.33}$ &  $-10.69 _{\scriptscriptstyle 0.26} ^{\scriptscriptstyle 0.91}$ &                                                             - &                                                             - &                                                             - &                                                             - &                                                             - &                                                             - \\
DES16X3cxn  &   $9.55 _{\scriptscriptstyle 0.08} ^{\scriptscriptstyle 0.57}$ &  $-0.32 _{\scriptscriptstyle 0.19} ^{\scriptscriptstyle 1.00}$ &   $-9.88 _{\scriptscriptstyle 0.11} ^{\scriptscriptstyle 0.43}$ &  $8.79 _{\scriptscriptstyle 0.36} ^{\scriptscriptstyle 0.16}$ &                                                             - &                                                             - &                                                             - &  $8.79 _{\scriptscriptstyle 0.36} ^{\scriptscriptstyle 0.16}$ &  $8.43 _{\scriptscriptstyle 0.27} ^{\scriptscriptstyle 0.18}$ \\
DES16X3ega  &   $9.94 _{\scriptscriptstyle 0.06} ^{\scriptscriptstyle 0.18}$ &   $0.30 _{\scriptscriptstyle 0.21} ^{\scriptscriptstyle 0.26}$ &   $-9.64 _{\scriptscriptstyle 0.15} ^{\scriptscriptstyle 0.08}$ &  $8.58 _{\scriptscriptstyle 0.18} ^{\scriptscriptstyle 0.18}$ &  $8.44 _{\scriptscriptstyle 0.20} ^{\scriptscriptstyle 0.17}$ &  $8.47 _{\scriptscriptstyle 0.09} ^{\scriptscriptstyle 0.08}$ &  $8.61 _{\scriptscriptstyle 0.06} ^{\scriptscriptstyle 0.07}$ &  $8.77 _{\scriptscriptstyle 0.08} ^{\scriptscriptstyle 0.06}$ &  $8.51 _{\scriptscriptstyle 0.08} ^{\scriptscriptstyle 0.08}$ \\
DES16X3erw  &   $9.89 _{\scriptscriptstyle 0.25} ^{\scriptscriptstyle 0.26}$ &   $0.94 _{\scriptscriptstyle 0.75} ^{\scriptscriptstyle 0.62}$ &   $-8.94 _{\scriptscriptstyle 0.51} ^{\scriptscriptstyle 0.36}$ &  $8.84 _{\scriptscriptstyle 0.13} ^{\scriptscriptstyle 0.09}$ &                                                             - &                                                             - &                                                             - &  $8.84 _{\scriptscriptstyle 0.13} ^{\scriptscriptstyle 0.09}$ &  $8.49 _{\scriptscriptstyle 0.13} ^{\scriptscriptstyle 0.10}$ \\
DES17C2hno  &   $9.56 _{\scriptscriptstyle 0.21} ^{\scriptscriptstyle 0.16}$ &  $-0.02 _{\scriptscriptstyle 0.44} ^{\scriptscriptstyle 0.33}$ &   $-9.58 _{\scriptscriptstyle 0.24} ^{\scriptscriptstyle 0.17}$ &  $8.78 _{\scriptscriptstyle 0.31} ^{\scriptscriptstyle 0.17}$ &                                                             - &                                                             - &                                                             - &  $8.78 _{\scriptscriptstyle 0.31} ^{\scriptscriptstyle 0.17}$ &  $8.43 _{\scriptscriptstyle 0.25} ^{\scriptscriptstyle 0.19}$ \\
DES17C3fwd  &  $10.07 _{\scriptscriptstyle 0.14} ^{\scriptscriptstyle 0.23}$ &   $0.84 _{\scriptscriptstyle 0.78} ^{\scriptscriptstyle 0.50}$ &   $-9.22 _{\scriptscriptstyle 0.64} ^{\scriptscriptstyle 0.27}$ &  $8.34 _{\scriptscriptstyle 0.14} ^{\scriptscriptstyle 0.12}$ &                                                             - &  $8.35 _{\scriptscriptstyle 0.17} ^{\scriptscriptstyle 0.18}$ &  $8.38 _{\scriptscriptstyle 0.12} ^{\scriptscriptstyle 0.08}$ &  $8.29 _{\scriptscriptstyle 0.10} ^{\scriptscriptstyle 0.11}$ &  $8.30 _{\scriptscriptstyle 0.09} ^{\scriptscriptstyle 0.10}$ \\
DES17C3gop  &   $9.90 _{\scriptscriptstyle 0.09} ^{\scriptscriptstyle 0.58}$ &   $0.19 _{\scriptscriptstyle 0.38} ^{\scriptscriptstyle 0.86}$ &   $-9.71 _{\scriptscriptstyle 0.28} ^{\scriptscriptstyle 0.28}$ &  $8.79 _{\scriptscriptstyle 0.48} ^{\scriptscriptstyle 0.21}$ &                                                             - &                                                             - &                                                             - &  $8.79 _{\scriptscriptstyle 0.48} ^{\scriptscriptstyle 0.21}$ &  $8.43 _{\scriptscriptstyle 0.28} ^{\scriptscriptstyle 0.24}$ \\
DES17S2fee  &  $11.20 _{\scriptscriptstyle 0.14} ^{\scriptscriptstyle 0.22}$ &   $0.28 _{\scriptscriptstyle 1.91} ^{\scriptscriptstyle 0.70}$ &  $-10.92 _{\scriptscriptstyle 1.77} ^{\scriptscriptstyle 0.48}$ &  $8.75 _{\scriptscriptstyle 0.47} ^{\scriptscriptstyle 0.52}$ &  $8.55 _{\scriptscriptstyle 0.48} ^{\scriptscriptstyle 0.42}$ &  $8.95 _{\scriptscriptstyle 0.45} ^{\scriptscriptstyle 0.47}$ &                                                             - &                                                             - &                                                             - \\
DES17X3cds  &   $9.36 _{\scriptscriptstyle 0.12} ^{\scriptscriptstyle 0.31}$ &  $-0.18 _{\scriptscriptstyle 0.25} ^{\scriptscriptstyle 0.83}$ &   $-9.53 _{\scriptscriptstyle 0.13} ^{\scriptscriptstyle 0.52}$ &  $8.69 _{\scriptscriptstyle 0.51} ^{\scriptscriptstyle 0.26}$ &                                                             - &                                                             - &                                                             - &  $8.69 _{\scriptscriptstyle 0.51} ^{\scriptscriptstyle 0.26}$ &  $8.33 _{\scriptscriptstyle 0.28} ^{\scriptscriptstyle 0.28}$ \\
DES17X3dxu  &  $10.49 _{\scriptscriptstyle 0.32} ^{\scriptscriptstyle 0.22}$ &   $1.02 _{\scriptscriptstyle 0.69} ^{\scriptscriptstyle 0.63}$ &   $-9.47 _{\scriptscriptstyle 0.37} ^{\scriptscriptstyle 0.40}$ &                                                             - &                                                             - &                                                             - &                                                             - &                                                             - &                                                             - \\
DES17X3hxi  &   $8.60 _{\scriptscriptstyle 0.11} ^{\scriptscriptstyle 0.00}$ &  $-0.37 _{\scriptscriptstyle 0.11} ^{\scriptscriptstyle 0.10}$ &   $-8.96 _{\scriptscriptstyle 0.00} ^{\scriptscriptstyle 0.10}$ &  $8.45 _{\scriptscriptstyle 0.45} ^{\scriptscriptstyle 0.26}$ &                                                             - &                                                             - &                                                             - &  $8.45 _{\scriptscriptstyle 0.45} ^{\scriptscriptstyle 0.26}$ &  $8.18 _{\scriptscriptstyle 0.18} ^{\scriptscriptstyle 0.18}$ \\

\bottomrule
\end{tabular}
\begin{tablenotes}
\item[a] Linear combination of the likelihoods for D16, PP04 N2, PP04 O3N2, KK04 R23.
\item[b] Weighted average of PP04 N2, PP04 O3N2, and KK04 R23, where N2 and R23 were converted to PP04 O3N2 via \citet{Kewley2008}.

\end{tablenotes}
\end{threeparttable}
\label{tab:derived}
\end{table*}

\subsection{Estimating metallicities \label{subsec:calc_Z}}

The most common method used to estimate the metallicity of galaxies is to use emission line ratios that have been calibrated using theoretical or empirical models in order to approximate the gas-phase oxygen abundance in the interstellar medium. Emission lines originate from regions of ionised gas, but there are a number of possible causes of this ionisation. Using the Baldwin-Phillips-Terlevich diagram (Fig.~\ref{fig:bpt}; \citealt{Baldwin1981}), we demonstrate that the emission line ratios measured in RET hosts are consistent with ionisation caused by star-formation as opposed to AGN. Only 12 of the 45 RET host spectra have the necessary lines to plot an \NII BPT diagram. This is also the case for the \SII~and \OI~versions of the diagram, and we find no evidence of AGN amongst the 12 hosts in those diagrams either.

Due to the low S/N of the spectra in this sample, we are constrained to a subset of metallicity diagnostics by the availability of only a handful of the strongest emission lines, namely \halpha, \hbeta, \OII 3727, \OIII 4959/5007, \NII 6548/6583, and \SII 6717/6731. Furthermore, for each host galaxy only a subset of these lines is detected - for example, \halpha, \NII~and \SII~are redshifted out of the spectral coverage at $z>0.3$, leaving only the oxygen and \hbeta~lines available and thus the R23 diagnostic, which is based off the relative strengths of the \OII and \OIII lines. For hosts at $z<0.3$ we are able to use the \OIII /\NII~(O3N2), \NII /\halpha~(N2), and \SII /\NII~(S2N2) line ratios. 

Due to the redshift range of our sample, and the limited wavelength coverage of the spectra ($3000-8000$~\AA), we are unable to use a single line ratio to estimate the oxygen abundances. We thus determine a set of indicators for which to calculate abundances. For the O3N2 and N2 indicators we use the calibration of \citet{Pettini2004} (PP04), and if \SII~is detected we derive an abundance using the S2N2 diagnostic of \citet{Dopita2016} (D16). For the R23 indicator, we use the calibration of \citet{Kobulnicky2004} (KK04). At abundances around $12 + \log \mathrm{(O/H)} \sim 8.4$, the R23 indicator becomes two-tailed, with a low and a high value of metallicity corresponding to a single R23 ratio. In cases where the lines are available, we break this degeneracy by cross-calibrating with the \NII/\OII~ratio \citep{Kewley2008}. In the cases where \NII~is not available, and there are no other diagnostics that can be used to inform the choice of branch, we use the host galaxy stellar mass to derive a crude metallicity estimate from the mass-metallicity relation (MZR) of \citet{Kewley2008} based upon the PP04 O3N2 diagnostic. For $12 + \log \mathrm{(O/H)}_{\mathrm{MZR}} < 8.4$ we chose the lower branch, while for higher MZR metallicities we choose the upper branch. We note that this is a rough estimation. If we leave the branch choice for those with no \NII~to be random, we find that the results are consistent to within uncertainties. 

The samples to which we compare metallicities span different redshift ranges, were observed with different equipment, and in many cases were compiled before certain (particularly the D16) diagnostics were devised. Therefore, in order to compare oxygen abundances between different samples we transform all abundances onto the PP04 O3N2 scale using the conversion factors given in \citet{Kewley2008}. This is not possible for the D16 diagnostic, so we discard it from the rest of our analysis, although for completeness we provide it for DES RET hosts where available. For samples that quoted multiple diagnostics, or for which sufficient line flux measurements were provided from which to calculate multiple diagnostics, we transform them all to the PP04 O3N2 scale. Following the prescription of \citet{Kruehler2015} we simultaneously minimise the oxygen abundance against the PDFs of the various different diagnostics scaled to PP04 O3N2, resulting in a final `best' PDF. We take $1\sigma$ uncertainties from the 16\textsuperscript{th} and 84\textsuperscript{th} percentiles of these PDFs, and for DES RETs display the results in Table \ref{tab:derived}.
\begin{figure}

\includegraphics[width=0.5\textwidth]{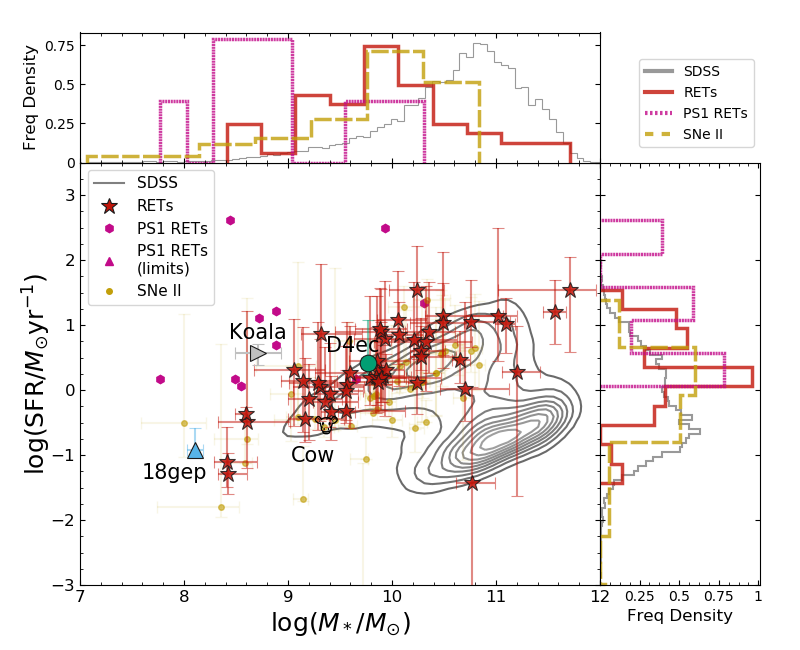}
\caption{The $M_* -$ SFR sequence of RET hosts, hosts of other rapid transients and SNe II, and SDSS field galaxies (grey contours). SFRs have not been corrected for redshift evolution. RET hosts lie slightly above the low-$z$ star-formation main sequence upper half of the SDSS contours, and systematically avoid passive galaxies (dense SDSS contours in the lower right).
\label{fig:sfms_sfr}}
\end{figure}

\section{Analysis}
\label{sec:analysis} 
A summary of the masses, sSFRs and metallicities for the hosts of DES RETs and various comparison samples is presented in Table \ref{tab:summary}. In the following sections, we compare the host galaxy properties of RETs to each sample in detail.

\begin{figure}
\includegraphics[width=0.5\textwidth]{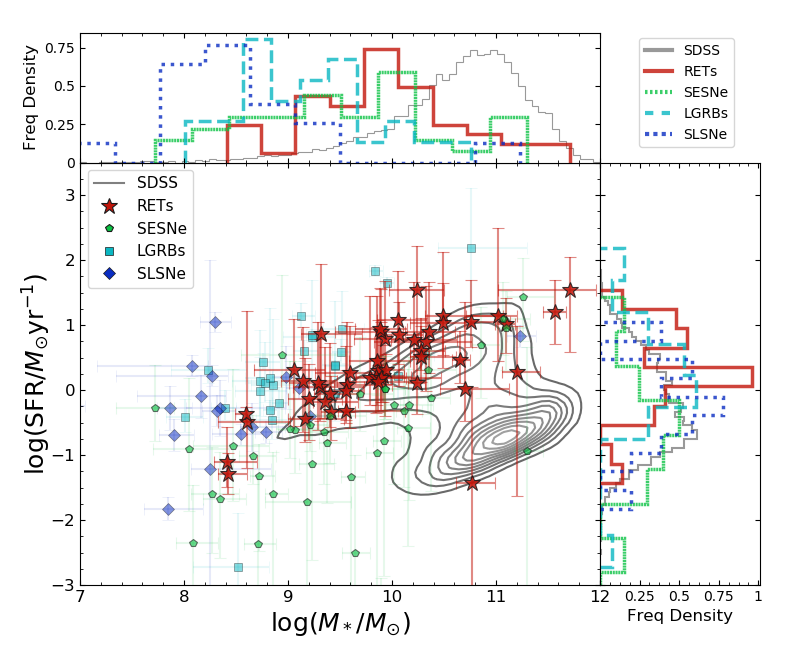}
\caption{Same as Fig.~\ref{fig:sfms_sfr}, but comparing RET hosts with hosts of SESNe, LGRBs and SLSNe, and SDSS field galaxies (grey contours).
\label{fig:sfms_sfr_other}}
\end{figure}

\subsection{Star formation rate \label{subsec:res_sfr}}
Figs.~\ref{fig:sfms_sfr} and \ref{fig:sfms_sfr_other}, split in two for clarity, show the `star formation main sequence' (SFMS) of RET host galaxies, as determined from photometric SED fitting along with that for the comparison samples and for the field galaxies of SDSS. RETs follow CCSNe, LGRBs and SLSNe in avoiding passive galaxies, evidence that RETs require the presence of star-formation and thus are linked to massive stars. One object (DES16C1cbd) lies among passive galaxies. The spectrum of this object is red in colour, but does exhibit \OII~emission indicative of recent star-formation activity consistent with the upper end of the sSFR error bar. The hosts of individual rapid transients Cow, Koala and SN2018gep are lower in mass than the majority of RETs. SN2018gep and Cow lie along the SFMS, while Koala sits in the starburst regime, which is not heavily populated by DES RETs. SNLS04D4ec is consistent with the peak of the DES RET mass and SFR distributions. 

\begin{figure}
\includegraphics[width=0.5\textwidth]{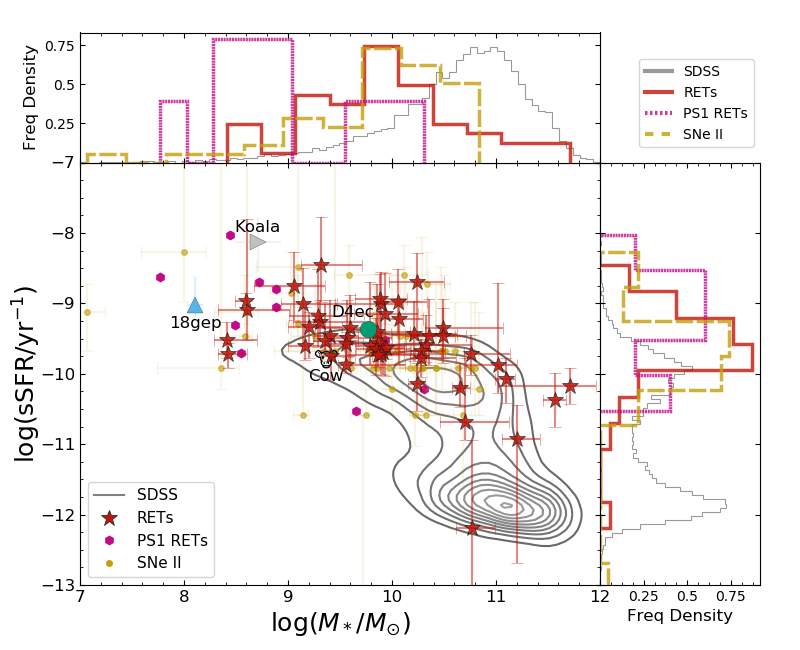}
\caption{The $M_* -$ sSFR sequence of RET hosts as well as local rapid transients and SNe II, along with the SDSS field galaxies.
\label{fig:sfms_ssfr}}
\end{figure}

\begin{figure}
\includegraphics[width=0.5\textwidth]{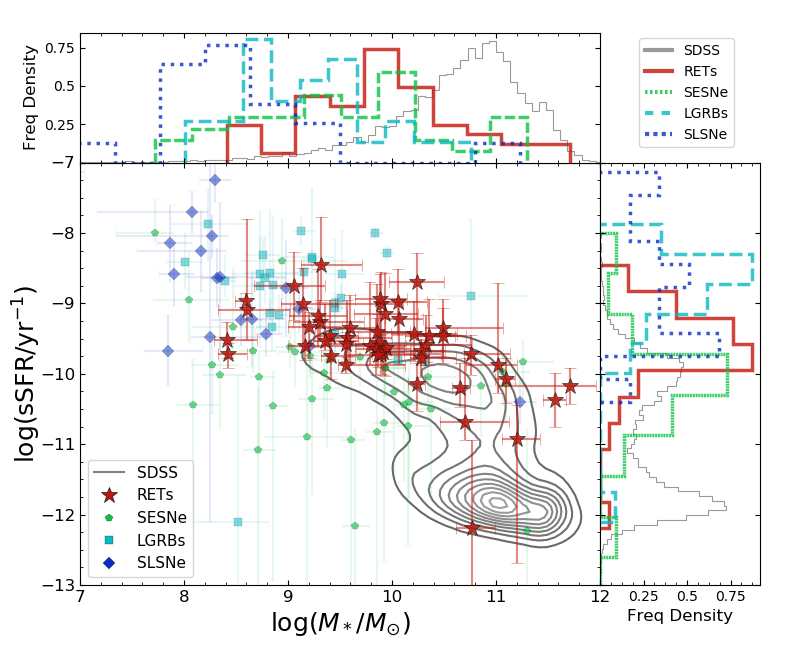}
\caption{Same as Fig.~\ref{fig:sfms_ssfr} but comparing RET hosts with hosts of SESN, LGRB, and SLSN hosts, along with the SDSS field galaxies.
\label{fig:sfms_ssfr_other}}
\end{figure}

Figs.~\ref{fig:sfms_ssfr} and \ref{fig:sfms_ssfr_other} are similar to Figs.~\ref{fig:sfms_sfr} and \ref{fig:sfms_ssfr_other}, except that here SFR has been normalised by stellar mass, and thus shows the specific star-formation rate (sSFR), which is a more representative measure of the star-forming efficiency. It is once again clear that RET hosts lie systematically above the majority of SDSS star-forming galaxies in terms of sSFR. Normalised by mass, it is here perhaps clearer to see that RET hosts lie at higher sSFR than CCSNe hosts, but not in the extremely star-forming environments of LGRBs or SLSNe.

\begin{figure}
\includegraphics[width=0.5\textwidth]{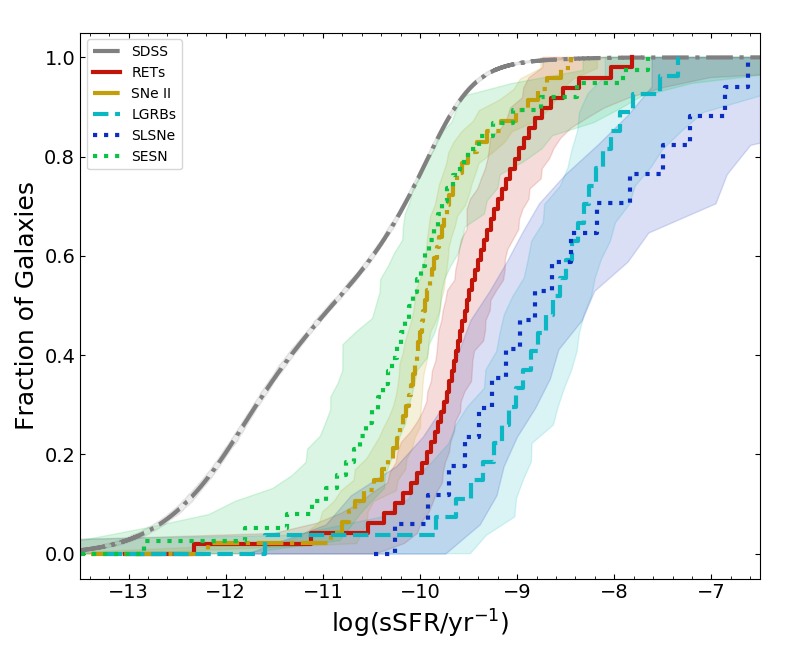}
\caption{Cumulative distributions of the sSFR of RET hosts, compared to CCSNe and the low-z SDSS sample. Uncertainties have been estimated via a bootstrap Monte-Carlo technique and include limits.
\label{fig:ssfr_cum}}
\end{figure}

We show the cumulative distribution of sSFR in Fig.~\ref{fig:ssfr_cum}. The RET hosts are clearly shifted to higher sSFRs than CCSNe. To statistically compare the host sSFR distribution of RETs with the other samples, we employ the method of \citetalias{Wiseman2020}. For each pair of samples, we model the PDFs as skewed normal distributions described by the parameters `loc' (location, identical to the mean for zero skewness), `scale' (spread, identical to the standard deviation for zero skewness)\footnote{See \citetalias{Wiseman2020} for a detailed description of the fitting procedure and the parameters describing the skewed normal distributions.}, and `skewness'. To impose priors on loc and scale, we combine the two samples and use normal priors centred on the combined mean and twice the combined standard deviation respectively, while for skewness the prior is a broad normal distribution centred on 0. We note that the loc parameter describes the location of the distribution (its relative position on the x-axis) and is not a mean, median, or mode. A highly-skewed distribution may have a loc that lies above almost the entire sample. A worked example as well as the results from the simultaneous fitting are displayed in Appendix \ref{app:b}.

The comparison shows RET hosts to be shifted to higher sSFRs than CCSNe. In 98\% of the posterior samples the RET sSFR distribution loc was at a higher value than SNe II, while the same was true 95\% of the time for RETs when comparing with SESNe. The mean difference is 0.6 dex. To test whether some of this could be attributed to the difference in redshift between the samples we apply an approximate redshift correction based on the parameterisation of the SFMS at different redshifts by \citet{Salim2007} and \citet{Noeske2007}, as has been done in other comparisons such as \citet{Taggart2019}. Transforming all host SFRs to their values at $z=0$ would result in the CCSN SFRs dropping by an average of 0.05 dex, while the RETs would decrease 0.35 dex, i.e. a difference of 0.30 dex, or half of the observed difference. The remaining 0.30 dex is thus consistent with being an intrinsic difference. 
RET host galaxies are significantly lower in sSFR than LGRBs. While the distributions are similar in shape, with a mean difference in scale of 0.09, the mean difference in loc is -1.16, with no overlap between the posterior distributions. The sSFR distribution of SLSNe hosts is much broader than the RETs, with a scale of 1.12, twice that of the RETs. They are also shifted to higher sSFRs than RETs, with the loc of the distribution on average 0.76 dex greater than RETs. The strong high-sSFR tail shows SLSNe occur in a different galaxy population to RETs.

\subsection{Metallicity \label{subsec:res_metallicity}}

\begin{figure}
\includegraphics[width=0.5\textwidth]{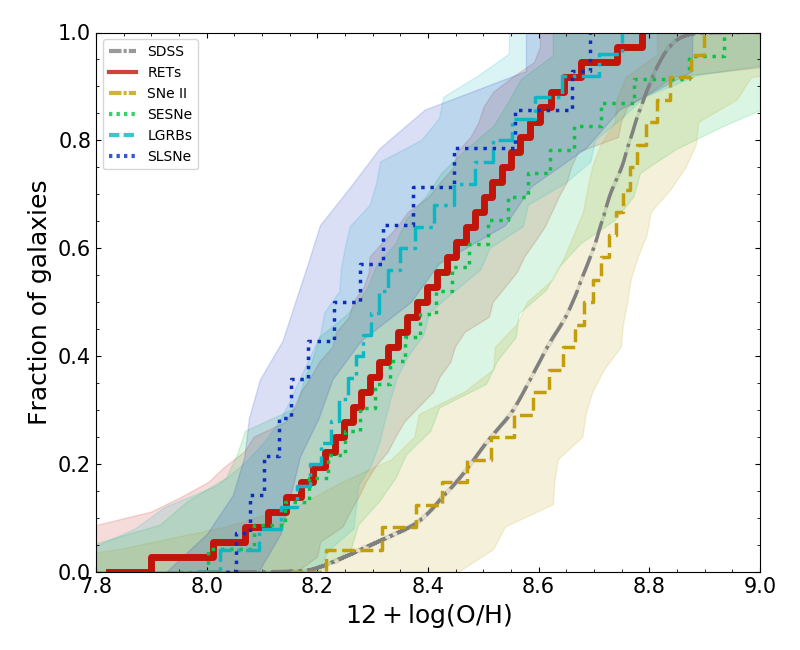}
\caption{Cumulative distributions of the gas-phase oxygen abundances of RET hosts, hosts of the comparison samples, and the SDSS sample. Uncertainties have been estimated via a bootstrap Monte-Carlo technique and include limits.
\label{fig:oh_cum}}
\end{figure}

In the Section \ref{subsec:res_sfr} we demonstrate that RETs occur in galaxies with systematically higher sSFR than CCSNe, to which one explanation is that they are related to more massive stars. A further property that could directly impact the composition of stellar populations harbouring potential RET progenitors is the metallicity. Using the gas-phase oxygen abundances calculated in Section \ref{subsec:calc_Z} as a proxy for metallicity, we compare the chemical state of RET host galaxies with CCSNe and star-forming field galaxies. The cumulative distributions of metallicity are displayed in Fig.~\ref{fig:oh_cum}, and show RET hosts to be inconsistent with SNe II and field galaxies. The RET curve lies at lower metallicity than those galaxies, and appears visually similar to the curves for SESNe. The metallicity distribution of SESNe is, however, quite broad \citep[e.g.][]{Anderson2010}, with different subclasses showing different trends (with SNe Ic host environments exhibiting higher metallicity than Ib, and IIb much lower). RETs occur, on average, in slightly more metal-rich environments than LGRBs and SLSNe.

We compare the metallicity distributions in the same way as the sSFRs, with the distribution fits shown in Appendix \ref{app:c}. The RET host metallicity distribution shows a broad peak, leading to two families of skewed-Gaussians that fit it well, one with a low-valued centre  ($12 + \log \mathrm{(O/H)} \sim 8.1$) and a positive skew, and the other with a higher-valued centre ($12 + \log \mathrm{(O/H)} \sim 8.6$). Comparing the DES RETs to the \citet{Stoll2013} SNe II shows the latter to be centred around 8.8, with the centre being greater than the RETs in 94\% of samples. We determine that the RET host metallicities are derived from a different population to the SNe II. On the other hand, simultaneous fits with SESNe show very similar distributions, including a smaller higher-metallicity peak, such that they are indistinguishable statistically. 

The median CDFs of LGRBs and SLSNe show divergence from the RETs, particularly at low metallicity. As a result, the locs of their fits are shifted compared to the RETs. In 67\% of samples, the RET sample had a higher loc than the LGRBS, with RETs also showing a broader distribution 67\% of the time. While these effects are not as significant as with the RETs - SNe II comparison, there is mild evidence that RETs are located in galaxies with higher metal content than LGRBs. The effect is more pronounced for SLSNe, where the RETs have a higher metallicity for the distribution peak 92\% of the time. The SLSN distribution is also more strongly skewed, with 89\% of the posterior distribution being more strongly skewed than the RETs. There is thus mild-to-strong evidence that RETs occur in more metal-rich environments than SLSNe. 

\begin{figure}
\includegraphics[width=0.5\textwidth]{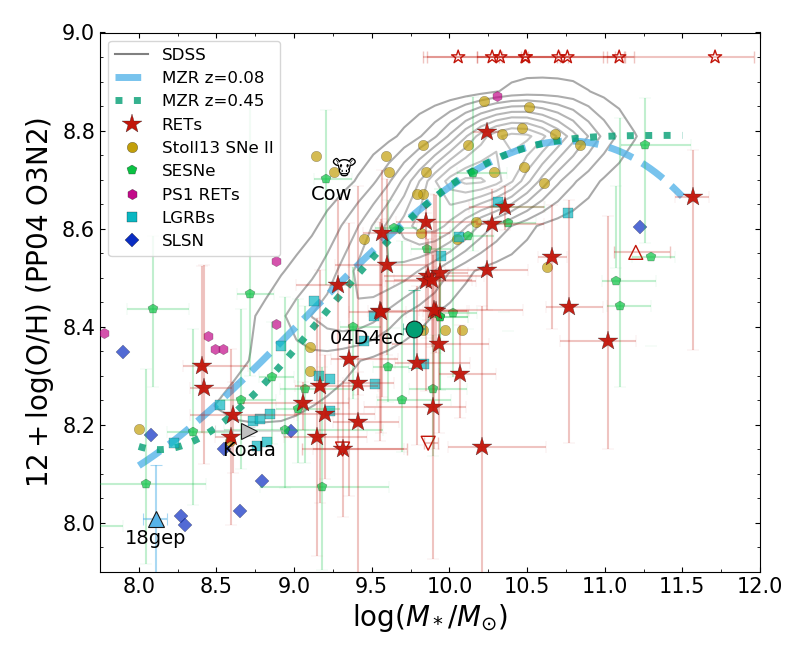}
\caption{The mass-metallicity relation (MZR) for RET host galaxies and comparison samples. Upward- and downward-pointing triangles reflect lower and upper limits respectively. The DES RETs with no metallicity measurement have been placed at the top of the figure for completeness. The dashed lines represent MZR parameterisations from \citet{Zahid2014}.
\label{fig:mzr}}
\end{figure}

In Fig.~\ref{fig:mzr} we show the MZR for the RET and comparison samples. The contours show the MZR for low-redshift ($\hat{z}=0.08$) star-forming galaixes from SDSS, adjusted to the PP04 O3N2 diagnostic. We use the MZR parameterisation \citet{Zahid2014} to show the best fit to the MZR for star-forming galaxies. The blue dashed line shows the fit to the low-z data, while the green dashed line corresponds to the MZR at $z=0.45$, the mean redshift of the RET host sample. The RET hosts lie systematically below the galaxy MZR fits as well as the bulk of the SDSS galaxies, meaning that for a given stellar mass they have a lower metallicity. They populate similar regions to SESNe, LGRBs and SLSNe but are clearly offset from the SNe II. SN2018gep, the Koala, and SNLS04D4ec appear in line with the RETs, while the Cow lies quite distinctly above the MZR, and is even outside the bulk of the local field galaxies.

\begin{figure}
\includegraphics[width=0.5\textwidth]{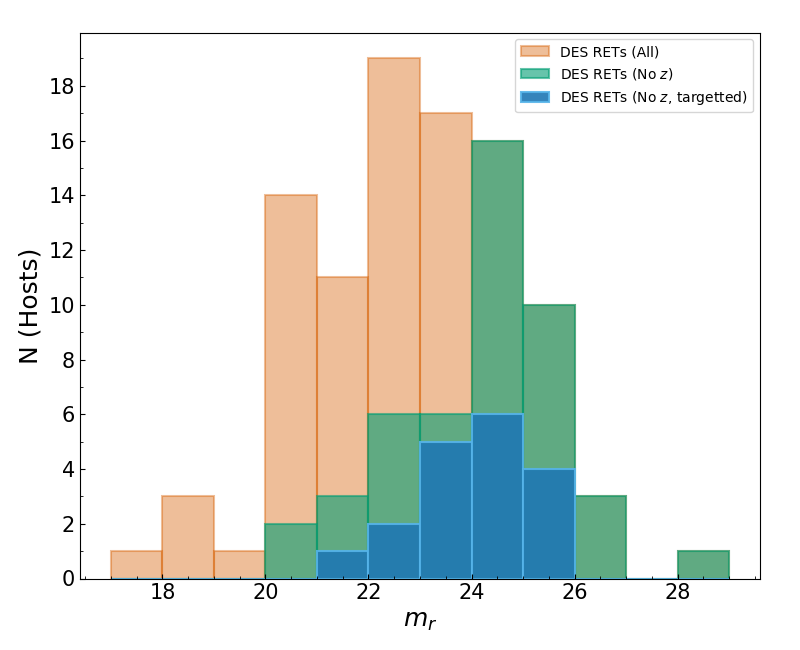}
\caption{Observer-frame $r$-band magnitude distribution for the host galaxies of RETs in DES. The orange histogram represents the 96/106 DES RETs for which a host was detected. The green histogram shows those that did not have a successful redshift measurement, while blue shows those with no redshift despite being targetted by OzDES.
\label{fig:mag_dist}}
\end{figure}

\begin{table}
    \centering
    \caption{Summary statistics for the samples compared in this work. Here we show the mean and standard deviation of the stellar mass, sSFR, and oxygen abundance, but note that the values for these parameters are not necessarily normally distributed for each sample. }
    \begin{tabular}{lccc}
    \toprule
   Sample & $\langle\log \left(M_*\right)\rangle$ &  $\langle\log \left(\mathrm{sSFR}\right)\rangle$ &$\langle12+\log\left(\mathrm{O/H}\right) \rangle$\\
   \midrule
   {} & $\left( M_{\odot}\right)$ & $\left(\mathrm{yr}^{-1} \right)$& \\
   \midrule
   DES RETs &$9.9 \pm 0.8$ & $-9.6 \pm 0.6 $ &$ 8.4 \pm 0.2$ \\
    PS1 RETs &$9.0 \pm 0.8$ & $-9.2 \pm 0.8 $ &$ 8.5 \pm 0.2$ \\
    SNe II &$9.8 \pm 0.8$ & $-9.8 \pm 0.7 $ &$ 8.6 \pm 0.2$ \\
    SESNe &$9.5 \pm 0.9$ & $-10.2 \pm 1.0 $ &$ 8.4 \pm 0.2$ \\
    LGRBs &$9.2 \pm 0.6$ & $-8.9 \pm 0.8 $ &$ 8.4 \pm 0.2$ \\
    SLSNe &$8.5 \pm 0.9$ & $-8.7 \pm 0.9 $ &$ 8.3 \pm 0.2$ \\
\bottomrule
    \end{tabular}
    
    \label{tab:summary}
\end{table}
\section{Discussion}

\subsection{Selection Biases \label{subsec:disc_bias}}
The properties presented in Section \ref{sec:analysis} are derived from a subset of the total sample of RETs. Of 106 objects, under half (52/106) have secure host galaxy redshifts. Three of these were obtained from transient spectra, for which we are unable to disentangle the host and transient contributions, and four were obtained by programmes for which we do not have access to the spectra. Of the remaining 45, it was possible to derive a metallicity or at least a limit for 40 host galaxies, while five exceeded the redshift range for the necessary lines to fall within the wavelength coverage of AAOmega. The observed metallicity distribution could have arisen if the galaxies without redshifts (and metallicities) are systematically higher in metallicity than those for which measurements were possible. For low SNR objects, redshifts are typically obtained from only two of the strongest lines (e.g. \halpha, \hbeta, \OIII, and \OII). It is likely that the redshifts were not obtained because the galaxies are physically smaller or are at higher redshift. However, galaxies with high metallicity have weaker \OIII~lines, meaning they are less likely to have a redshift detection compared to less enriched galaxies with the same mass and redshift. Future, deeper spectral observation programmes as well as large, complete low-redshift samples are necessary to reduce this possible bias.

\begin{figure}
\includegraphics[width=0.5\textwidth]{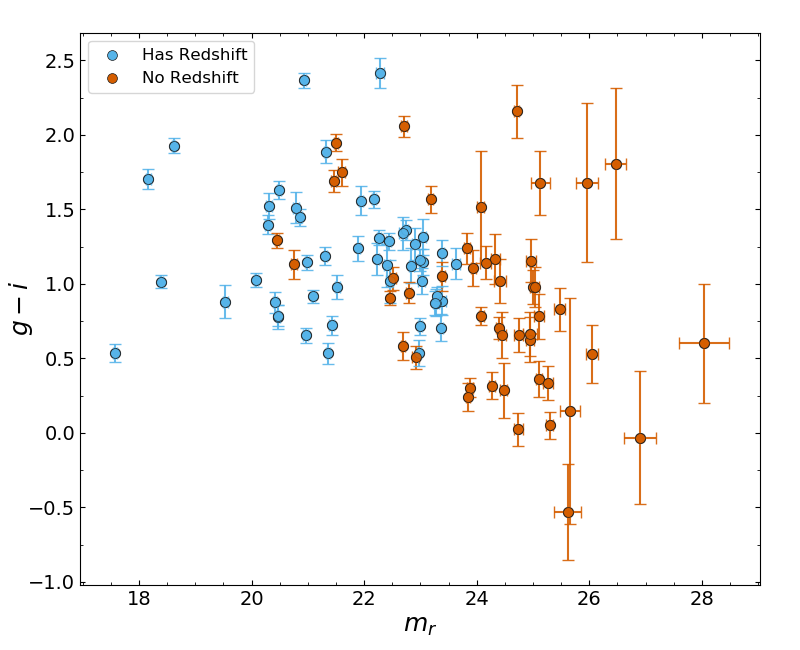}
\caption{The colour-magnitude distribution of RET hosts with (cyan) and without (orange) redshifts. There is an excess of objects with blue colours that do not have redshift measurements.
\label{fig:g-i}}
\end{figure}

Another possibility is that the hosts without a redshift are mostly non star-forming, passive galaxies, for which a redshift is typically harder to obtain than for emission-line galaxies \citep{Yuan2015,Childress2017,Lidman2020}. To test this possibility, we examined the RETs that do not have a host galaxy redshift. Table \ref{tab:z_cuts} shows the numbers of RETs that failed various stages of the redshifting process, and is summarised in Fig.~\ref{fig:mag_dist}. Of the 57 objects without a redshift, 47 of them have host galaxies detected in the SN Deep coadds of \citetalias{Wiseman2020}. Of more significance is that only 40 have host galaxies in the SVA1 catalogues which were used for targeting during the OzDES campaign. The other, `hostless', objects are either transients that are located remotely from a galaxy that was detected, or are hosted by a galaxy that was not detected. Non-detected hosts are either intrinsically faint and thus low in mass, situated at high redshift, or both. Neither are expected to be systematically higher in metallicity than the detected hosts. Similarly, a further 22 hosts were detected but not targeted by OzDES, due to being too faint to pass the selection criteria ($m_r < 24.5$), leaving 18 that were targetted but no redshift was found. The resulting redshift completeness of targeted objects is 71\% (83\% for objects brighter than $m_r = 24$~mag), which is in line with the average for OzDES \citep{Lidman2020}.
In Fig.~\ref{fig:g-i} we show the observer-frame $r$-band magnitudes and $g-i$ colours for all RET hosts that were detected. The 47 objects with detected hosts but no redshift lie at fainter magnitudes, and appear to extend to bluer colours than those with secure redshifts. This is contrary to the hypothesis that they are high-redshift and/or passive hosts, but instead are low-mass, star-forming galaxies whose line fluxes were not strong enough to be detected. We thus conclude that the results presented from the subset of hosts with measured redshifts are at the very least representative of the star-forming nature of the population of RET hosts.

\begin{table}
    \centering
    \begin{tabular}{l|l}
         Cut &  Number of remaining objects \\
        \toprule
        All RETs & 106 \\
        No redshift   & 57\\
        Has host in SN Deep & 47\\
        Has host in SVA1 & 40 \\
        Targetted by OzDES & 18 \\
        \bottomrule
    \end{tabular}
    \caption{Numbers of RETs passing various cuts relating to redshift targetting and completeness. Each row is a subset of the row above.}
    \label{tab:z_cuts}
\end{table}

\subsection{Origin of RETs \label{subsec:disc_origin}}

The sample of DES RETs shows a preference for low-metallicity, strongly star-forming host environments. The PDF of their metallicities displays a strong similarity to the hosts of SESNe, as well as LGRBs. There is a clear difference to the PDF of SNe II, which follow SDSS field galaxies. The preference for low-metallicity systems is not as strong as for LGRBs or SLSNe, but the highest metallicities found in all three samples are very similar at around solar metallicity. This result is suggestive of a stripped-envelope, massive-star origin for RETs. 
The population of RET hosts lies, on average, between CCSNe and LGRBs/SLSNe in terms of both star formation and metallicity. A loose correlation exists between the luminosity and rarity of events, and the host galaxy conditions required for their formation $-$ on average, rarer events occur in more extreme environments. The approximate rate of RETs ($\geq 10^{-6} \mathrm{Mpc}^{-3} \mathrm{yr}^{-1}$), \citealt{Drout2014}, \citetalias{Pursiainen2018}, \citealt{Tampo2020}) is $\sim1\%$ of the CCSN rate \citep{Li2011,Horiuchi2011,Strolger2015}, which itself is divided into the more common SNe II and sub-dominant SESNe \citep{Kelly2012,Frohmaier2020}. At $\sim1\%$ of the CCSN rate, RETs are more common than SLSNe ($\sim0.01 - 0.05\%$ of CCSNe; \citealt{McCrum2015,Prajs2017,Frohmaier2020}) and LGRBs (intrinsically $\sim0.08\%$ when accounting for beaming; \citealt{Graham2016}). These figures place the rate of DES RETs between extreme objects (SLSNe, LGRBs) and more common SNe (SNe II, SESNe) in terms of rate, matching the location of RET hosts in the various host galaxy parameter spaces presented in Section \ref{sec:analysis}. While stressing rates are uncertain and host galaxy parameters span wide ranges for all transients, they are both linked to the respective transients' progenitor channels. Based upon both indicators, it is reasonable to infer that RETs are linked to very massive stars, potentially stripped of their envelopes, and possibly sharing some of the extreme properties of SLSNe or LGRBs. This hypothesis also suggests that RETs are an intermediate and/or precursory step, whereby the initial collapse of the star occurs leading to shock breakout and subsequent cooling, but conditions are not highly tuned enough for a LGRB or SLSN and the respective central engine does not form. 

\subsection{Correlations between lightcurve and host galaxy properties \label{subsec:disc_correlations}}
\begin{figure}
\includegraphics[width=0.5\textwidth]{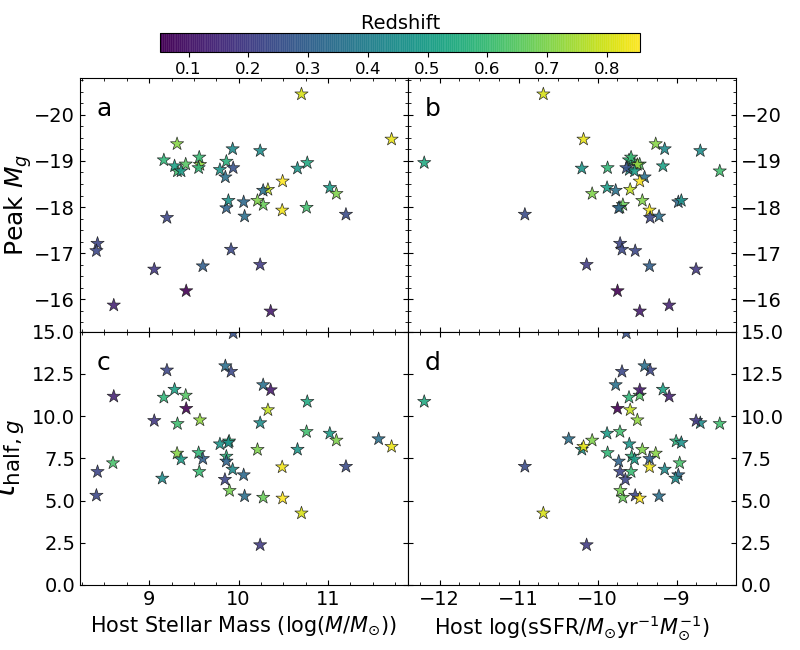}
\caption{RET lightcurve properties as a function of host galaxy measurements.
\label{fig:ret_v_host}}
\end{figure}

Many classes of transients show trends between properties intrinsic to the objects themselves and their host galaxies. For example, SNe Ia lightcurves appear to be broader in less massive galaxies with higher sSFRs \citep{Sullivan2006,Neill2009,Howell2009,Sullivan2010,Roman2018,Kelsey2020}, while SLSN lightcurves that have been fit with a magnetar model show a tentative relationship between the magnetar spin period and host galaxy metallicity \citep{Chen2016a}. In Fig.~\ref{fig:ret_v_host} we show the RET peak magnitude (upper panels) and lightcurve width parameterised as $t_{\mathrm{half}}$, the time the lightcurve is above half the peak brightness (lower panels), and their correlation with host galaxy stellar mass (left-hand panels) and sSFR (right-hand panels). The decline rates have been converted to the rest-frame of the transients, while the peak magnitudes have been k-corrected assuming a blackbody SED. There is no correlation between decline rate and either stellar mass or sSFR, while there are hints of a trend between peak magnitude and both mass and sSFR. These apparent trends are driven by the more extreme hosts (the three with $\log\left(M_*/M_{\odot}\right) < 9$ and one with very high mass/low sSFR). Assuming that these points are not outliers, the trends are still likely driven by selection effects. At higher redshifts, only the brighter transients are recovered by the survey and our selection method, while at those high redshifts only the more massive galaxies are detected. This effect can be seen in Fig.~\ref{fig:ret_v_host}a, with redshift increasing from the lower left to the upper right, while the same is true from the upper left to lower right in Fig.~\ref{fig:ret_v_host}b. It is hoped that a more complete, volume-limited sample of RETs will be obtained by The Rubin Observatory Legacy Survey of Space and Time (LSST) in order to reveal any underlying relationships.

\label{sec:disc}

\subsection{Comparison with individual RETs \label{subsec:disc_lowz}}

The nearby transient AT2018cow has drawn many comparisons to the cosmological RETs from DES and PS1 \citep[e.g.][]{Perley2019,Margutti2019,Fox2019,Mohan2020} due to its rapid evolution and blue colour. AT2018cow displayed a contracting photosphere as well as evidence for central-engine power alongside an unusual spectrum that showed similarities to broad-lined SNe Ic (SN Ic-bl) at early stages \citep[e.g.][]{Xu2018,Izzo2018}, developing to something entirely different at later epochs \citep{Perley2019} with hints of similarities to interacting SNe Ibn \citep{Fox2019}. There have been several suggestions that AT2018cow is indeed an analogue of the high-z RETs. The host galaxy of AT2018cow is to be moderately star forming and lies very close to the centre of the SFMS (\citealt{Lyman2020}; Figs. \ref{fig:sfms_sfr},\ref{fig:sfms_ssfr}), along with many of the DES RET hosts. However, the host lies somewhat above the fiducial MZR in Fig.~\ref{fig:mzr}, suggesting that it has an unusually high metallicity for its stellar mass. While consistent with SNe II, this is in contrast to the DES RET hosts which are systematically less enriched for a given stellar mass. 

Other local rapid transients include SN2018gep \citep{Ho2019}, a spectroscopically classified SN Ic-bl with a rapid rise. The host of SN2018gep appears more similar to the DES RET sample, lying in the same $M_*$-SFR and $M*$-sSFR plane, as well as lying below the MZR. While the SN2018gep host is lower in stellar mass than any DES RET $\left(\log \left(M/M_{\odot}\right) =8.11\right)$, galaxies of that mass are unlikely to have been detected at the redshifts of the DES RETs \citep{Wiseman2020}. The authors' conclusion that SN2018gep is related to a shock-breakout of a massive, stripped-envelope star is similar to that posited in Section \ref{subsec:disc_origin}. 

The rapidly evolving lightcurve of ZTF18abvkwla (``the Koala'') has been attributed to shock interaction, while radio emission can be explained by a collimated jet. The host of the Koala is a low metallicity starburst more typical of LGRBs and SLSNe, and places this transient at the very extreme end of the DES RET host population. While we note that the \citet{Ho2020} study made multiple non-detections of radio emission from the DES RETs, these were taken at very late epochs ($\geq 1~\mathrm{year}$), so the presence of jets in the early evolution is not ruled out. Similarly, we cannot rule out that the Koala comes from the same population of transients as the DES RETs.

SN2018kzr \citep{McBrien2019} is one of the most rapidly declining transients ever discovered, with spectral signatures similar to SNe Ic. While host galaxy properties are not derived, the authors of that paper refer to narrow emission from the host galaxy, along with an apparently small, blue, star-forming host and is thus consistent with the DES RETs.


\section{Conclusions}
\label{sec:conc}
By analysing the host galaxies of 49 rapidly evolving transients (RETs) discovered in the Dark Energy Survey, we have been able to place constraints on the nature of these as-yet unexplained phenomena. We conclude that RETs are strongly linked to massive stars, due to their hosts all exhibiting signatures of star formation. They likely originate from stars more massive, on average, than than those that cause SNe II, and perhaps all SESNe, as they occur in galaxies with higher sSFR. RET hosts are significantly lower in metallicity than SN II hosts, and marginally lower than SESN hosts, suggesting some reliance on rotational energy or other metallicity-dependent effects.
Of the RET analogues discovered in modern large-area, high-cadence surveys, ZTF18abvkwla shares the most similar host galaxy characteristics with the DES RET population. SN2018gep appears in a galaxy too faint to have been detected by DES-SN at the redshift of most of the DES RETs, while the host of AT2018cow is higher in metallicity.

While current surveys such as ZTF \citep{Bellm2019}, GOTO \citep{Dyer2018}, and BlackGEM \citep{Bloemen2016} are well equipped to find low-redshift RETs, a sample similar to that presented here will likely not be collected until LSST comes online. With several hundreds of objects, detailed studies of RETs and their hosts will be possible in a systematic and more complete manner as has been achieved with LGRBs and SLSNe.

\section*{Acknowledgements}

We acknowledge support from STFC grant ST/R000506/1. MSm, MSu, and CPG acknowledge support from EU/FP7-ERC grant 615929. LG was funded by the European Union's Horizon 2020 research and innovation programme under the Marie Sk\l{}odowska-Curie grant agreement No. 839090. This work has been partially supported by the Spanish grant PGC2018-095317-B-C21 within the European Funds for Regional Development (FEDER). L.K. was supported by the Science and Technology Facilities Council (grant number ST/P006760/1) through the DISCnet Centre for Doctoral Training.

This paper makes use of observations taken using the Anglo-Australian Telescope under programs ATAC A/2013B/12 and NOAO 2013B-0317.

This research used resources of the National Energy Research Scientific Computing Center (NERSC), a U.S. Department of Energy Office of Science User Facility operated under Contract No. DE-AC02-05CH11231.

Funding for the DES Projects has been provided by the U.S. Department of Energy, the U.S. National Science Foundation, the Ministry of Science and Education of Spain, 
the Science and Technology Facilities Council of the United Kingdom, the Higher Education Funding Council for England, the National Center for Supercomputing 
Applications at the University of Illinois at Urbana-Champaign, the Kavli Institute of Cosmological Physics at the University of Chicago, 
the Center for Cosmology and Astro-Particle Physics at the Ohio State University,
the Mitchell Institute for Fundamental Physics and Astronomy at Texas A\&M University, Financiadora de Estudos e Projetos, 
Funda{\c c}{\~a}o Carlos Chagas Filho de Amparo {\`a} Pesquisa do Estado do Rio de Janeiro, Conselho Nacional de Desenvolvimento Cient{\'i}fico e Tecnol{\'o}gico and 
the Minist{\'e}rio da Ci{\^e}ncia, Tecnologia e Inova{\c c}{\~a}o, the Deutsche Forschungsgemeinschaft and the Collaborating Institutions in the Dark Energy Survey. 

The Collaborating Institutions are Argonne National Laboratory, the University of California at Santa Cruz, the University of Cambridge, Centro de Investigaciones Energ{\'e}ticas, 
Medioambientales y Tecnol{\'o}gicas-Madrid, the University of Chicago, University College London, the DES-Brazil Consortium, the University of Edinburgh, 
the Eidgen{\"o}ssische Technische Hochschule (ETH) Z{\"u}rich, 
Fermi National Accelerator Laboratory, the University of Illinois at Urbana-Champaign, the Institut de Ci{\`e}ncies de l'Espai (IEEC/CSIC), 
the Institut de F{\'i}sica d'Altes Energies, Lawrence Berkeley National Laboratory, the Ludwig-Maximilians Universit{\"a}t M{\"u}nchen and the associated Excellence Cluster Universe, 
the University of Michigan, the National Optical Astronomy Observatory, the University of Nottingham, The Ohio State University, the University of Pennsylvania, the University of Portsmouth, 
SLAC National Accelerator Laboratory, Stanford University, the University of Sussex, Texas A\&M University, and the OzDES Membership Consortium.

Based in part on observations at Cerro Tololo Inter-American Observatory, National Optical Astronomy Observatory, which is operated by the Association of 
Universities for Research in Astronomy (AURA) under a cooperative agreement with the National Science Foundation.

The DES data management system is supported by the National Science Foundation under Grant Numbers AST-1138766 and AST-1536171.
The DES participants from Spanish institutions are partially supported by MINECO under grants AYA2015-71825, ESP2015-66861, FPA2015-68048, SEV-2016-0588, SEV-2016-0597, and MDM-2015-0509, 
some of which include ERDF funds from the European Union. IFAE is partially funded by the CERCA program of the Generalitat de Catalunya.
Research leading to these results has received funding from the European Research
Council under the European Union's Seventh Framework Program (FP7/2007-2013) including ERC grant agreements 240672, 291329, and 306478.
We  acknowledge support from the Brazilian Instituto Nacional de Ci\^encia
e Tecnologia (INCT) e-Universe (CNPq grant 465376/2014-2).

This manuscript has been authored by Fermi Research Alliance, LLC under Contract No. DE-AC02-07CH11359 with the U.S. Department of Energy, Office of Science, Office of High Energy Physics.

This work makes extensive use of Astropy,\footnote{http://www.astropy.org} a community-developed core Python package for Astronomy \citep{AstropyCollaboration2013,AstropyCollaboration2018}, Pandas \citep{Mckinney2010}, and matplotlib \citep{Hunter2007}.




\bibliographystyle{mnras}
\bibliography{PhilMendeley} 



\newpage
\appendix
\onecolumn
\section{Spectral line fluxes}
Table \ref{tab:fluxes} presents the line fluxes for all DES RET hosts for which spectra were available. Spectra are available from the public OzDES DR2 at \url{https://docs.datacentral.org.au/ozdes/overview/dr2/}.

\begin{table*}

\caption{Emission line fluxes for DES RET host galaxies. Values are given in units of erg s$^{-1}$ cm$^{-2}$ \AA$^{-1}$, and have been corrected for Milky Way reddening using \citet{Schlegel1998} assuming a \citet{Cardelli1989} reddening law with $R_V = 3.1$, but have not been corrected for intrinsic host galaxy reddening.\label{tab:fluxes}}
\setlength{\tabcolsep}{3pt}
\begin{tabular}{lccccccccccc}
\toprule
{} &              \OII 3727 &             \OIII 4960 &              \OIII 5007 &             \NII 6549 &             \NII 6585 &             \SII 6717 &             \SII 6731 &           \hdelta &            \hgamma &             \hbeta &            \halpha \\
\midrule
DES13X3gms  &   $1.2 \pm 14.8$ &  $0.3 \pm 1.8$ &   $0.9 \pm 1.8$ &              - &               - &              - &              - &  $0.0 \pm 1.9$ &  $15.1 \pm 2.0$ &   $0.7 \pm 1.5$ &               - \\
DES13C1tgd  &    $1.6 \pm 1.5$ &  $0.0 \pm 0.5$ &   $0.0 \pm 0.5$ &  $0.7 \pm 0.6$ &   $2.2 \pm 0.6$ &  $1.5 \pm 0.3$ &  $1.0 \pm 0.4$ &  $0.4 \pm 0.5$ &   $1.1 \pm 0.5$ &   $0.3 \pm 0.5$ &   $5.8 \pm 0.8$ \\
DES13S2wxf  &  $32.7 \pm 10.5$ &  $1.3 \pm 1.1$ &   $4.1 \pm 1.1$ &              - &               - &              - &              - &  $1.9 \pm 1.2$ &   $2.1 \pm 1.0$ &   $1.8 \pm 1.1$ &               - \\
DES13X1hav  &    $3.0 \pm 1.2$ &  $1.0 \pm 0.6$ &   $3.0 \pm 0.6$ &              - &               - &              - &              - &  $0.0 \pm 0.3$ &   $0.5 \pm 0.5$ &   $0.9 \pm 0.3$ &               - \\
DES13X3nyg  &    $2.6 \pm 0.6$ &  $0.1 \pm 0.4$ &   $0.3 \pm 0.4$ &              - &               - &              - &              - &  $0.1 \pm 0.2$ &   $0.0 \pm 0.2$ &   $0.1 \pm 0.1$ &               - \\
DES13X3gmd  &    $2.7 \pm 0.7$ &              - &               - &              - &               - &              - &              - &  $0.1 \pm 0.4$ &   $0.3 \pm 0.4$ &   $1.3 \pm 1.0$ &               - \\
DES13C3bcok &    $3.6 \pm 3.8$ &  $0.7 \pm 1.2$ &   $2.2 \pm 1.2$ &              - &               - &              - &              - &  $0.0 \pm 1.2$ &   $2.0 \pm 1.4$ &   $2.3 \pm 1.0$ &  $10.8 \pm 3.0$ \\
DES13X2wvv  &   $16.4 \pm 2.7$ &  $2.9 \pm 1.4$ &   $8.8 \pm 1.4$ &              - &               - &              - &              - &  $1.1 \pm 1.3$ &   $0.6 \pm 1.2$ &   $4.3 \pm 1.1$ &               - \\
DES14X1bnh  &    $8.3 \pm 0.8$ &              - &               - &              - &               - &              - &              - &  $0.1 \pm 0.3$ &   $1.0 \pm 0.3$ &               - &               - \\
DES15S1fli  &   $15.0 \pm 1.6$ &  $1.3 \pm 1.0$ &   $3.9 \pm 1.0$ &              - &               - &              - &              - &  $1.4 \pm 0.8$ &   $2.2 \pm 0.6$ &   $4.3 \pm 0.5$ &               - \\
DES15S1fll  &   $10.3 \pm 2.5$ &  $5.0 \pm 2.0$ &  $15.2 \pm 2.0$ &  $0.5 \pm 1.1$ &   $1.4 \pm 1.1$ &  $0.8 \pm 0.8$ &  $1.5 \pm 0.8$ &  $0.6 \pm 1.1$ &   $2.5 \pm 1.3$ &   $4.2 \pm 1.5$ &  $15.4 \pm 1.4$ \\
DES14X3pkl  &    $2.5 \pm 1.2$ &  $0.8 \pm 0.4$ &   $2.3 \pm 0.4$ &  $0.2 \pm 0.2$ &   $0.5 \pm 0.2$ &  $0.8 \pm 0.2$ &  $0.4 \pm 0.2$ &  $0.4 \pm 0.5$ &   $0.6 \pm 0.6$ &   $0.0 \pm 0.7$ &   $2.3 \pm 0.3$ \\
DES13X3npb  &   $15.9 \pm 6.6$ &  $0.8 \pm 0.8$ &   $2.4 \pm 0.8$ &              - &               - &              - &              - &  $2.0 \pm 0.7$ &   $2.6 \pm 0.4$ &   $3.3 \pm 0.8$ &               - \\
DES15X2ead  &  $31.4 \pm 23.5$ &  $1.6 \pm 3.0$ &   $4.8 \pm 3.0$ &  $1.1 \pm 1.6$ &   $3.5 \pm 1.6$ &  $5.3 \pm 1.7$ &  $3.2 \pm 1.9$ &  $9.6 \pm 7.9$ &   $4.5 \pm 6.0$ &   $5.4 \pm 3.1$ &  $16.7 \pm 1.9$ \\
DES14S2plb  &   $54.5 \pm 7.2$ &  $4.5 \pm 2.1$ &  $13.5 \pm 2.1$ &  $5.6 \pm 1.8$ &  $16.9 \pm 1.8$ &  $9.5 \pm 0.9$ &  $7.1 \pm 0.7$ &  $6.5 \pm 2.0$ &   $7.9 \pm 1.7$ &  $19.3 \pm 1.5$ &  $53.4 \pm 1.7$ \\
DES14S2pli  &    $7.7 \pm 1.1$ &  $0.4 \pm 0.4$ &   $1.2 \pm 0.4$ &              - &               - &              - &              - &  $0.0 \pm 0.6$ &   $1.1 \pm 0.5$ &   $2.4 \pm 0.3$ &               - \\
DES14C3tnz  &    $4.2 \pm 0.9$ &  $0.4 \pm 0.3$ &   $1.3 \pm 0.3$ &              - &               - &              - &              - &  $0.0 \pm 0.3$ &   $0.4 \pm 0.3$ &   $0.4 \pm 0.5$ &               - \\
DES15X3mxf  &    $4.8 \pm 0.7$ &  $0.4 \pm 0.3$ &   $1.2 \pm 0.3$ &              - &               - &              - &              - &  $0.6 \pm 0.5$ &   $0.7 \pm 0.4$ &   $1.1 \pm 0.3$ &               - \\
DES15C3lpq  &    $3.9 \pm 0.8$ &  $0.5 \pm 0.3$ &   $1.6 \pm 0.3$ &              - &               - &              - &              - &  $0.5 \pm 0.3$ &   $0.2 \pm 0.2$ &   $0.6 \pm 0.2$ &               - \\
DES15C3nat  &    $3.9 \pm 0.6$ &              - &               - &              - &               - &              - &              - &  $0.2 \pm 0.1$ &   $0.2 \pm 0.1$ &               - &               - \\
DES15C3mgq  &    $4.1 \pm 2.5$ &  $1.1 \pm 0.5$ &   $3.2 \pm 0.5$ &  $0.1 \pm 0.2$ &   $0.2 \pm 0.2$ &  $0.2 \pm 0.2$ &  $0.0 \pm 0.4$ &  $0.6 \pm 0.8$ &   $0.4 \pm 0.7$ &   $1.2 \pm 0.5$ &   $1.4 \pm 0.2$ \\
DES15E2nqh  &    $0.7 \pm 1.0$ &  $0.9 \pm 0.5$ &   $2.8 \pm 0.5$ &              - &               - &              - &              - &  $0.4 \pm 0.4$ &   $0.9 \pm 0.5$ &   $0.8 \pm 0.5$ &               - \\
DES15C3opk  &    $1.6 \pm 0.8$ &  $0.5 \pm 0.3$ &   $1.4 \pm 0.3$ &              - &               - &              - &              - &  $0.1 \pm 0.3$ &   $0.3 \pm 0.2$ &   $0.4 \pm 0.3$ &               - \\
DES15C3opp  &    $2.5 \pm 0.7$ &  $0.2 \pm 0.2$ &   $0.5 \pm 0.2$ &              - &               - &              - &              - &  $0.2 \pm 0.4$ &   $0.5 \pm 0.3$ &   $0.4 \pm 0.1$ &               - \\
DES16E2pv   &    $2.5 \pm 0.6$ &  $0.7 \pm 1.0$ &   $2.2 \pm 1.0$ &              - &               - &              - &              - &  $0.0 \pm 0.4$ &   $0.7 \pm 0.5$ &   $1.3 \pm 0.6$ &               - \\
DES16S1bbp  &   $21.9 \pm 3.5$ &  $3.8 \pm 0.9$ &  $11.4 \pm 0.9$ &  $0.3 \pm 0.4$ &   $1.0 \pm 0.4$ &  $1.5 \pm 0.7$ &  $0.9 \pm 0.3$ &  $1.4 \pm 1.1$ &   $3.4 \pm 0.9$ &   $6.4 \pm 1.0$ &  $11.1 \pm 0.6$ \\
DES16X3cxn  &    $2.3 \pm 0.5$ &  $0.3 \pm 0.4$ &   $0.9 \pm 0.4$ &              - &               - &              - &              - &  $0.0 \pm 0.3$ &   $0.2 \pm 0.2$ &   $0.6 \pm 0.2$ &               - \\
DES16C1cbd  &    $4.4 \pm 1.8$ &  $0.4 \pm 0.8$ &   $1.2 \pm 0.8$ &              - &               - &              - &              - &  $0.4 \pm 0.4$ &   $0.4 \pm 0.4$ &   $1.1 \pm 0.4$ &               - \\
DES16C3axz  &   $26.8 \pm 2.6$ &  $2.3 \pm 0.8$ &   $7.0 \pm 0.8$ &  $2.6 \pm 0.6$ &   $7.7 \pm 0.6$ &  $5.4 \pm 0.5$ &  $4.0 \pm 0.5$ &  $2.7 \pm 0.9$ &   $5.6 \pm 0.8$ &   $9.9 \pm 0.8$ &  $29.1 \pm 0.7$ \\
DES16X3erw  &    $5.9 \pm 0.9$ &  $0.8 \pm 0.6$ &   $2.6 \pm 0.6$ &              - &               - &              - &              - &  $0.6 \pm 0.4$ &   $0.4 \pm 0.4$ &   $1.9 \pm 0.3$ &               - \\
DES16C3gin  &    $5.8 \pm 1.2$ &  $0.5 \pm 0.4$ &   $1.6 \pm 0.4$ &              - &               - &              - &              - &  $0.5 \pm 0.6$ &   $0.0 \pm 0.6$ &   $1.6 \pm 0.4$ &   $2.5 \pm 0.4$ \\
DES16S1dxu  &   $29.2 \pm 3.6$ &  $6.4 \pm 1.9$ &  $19.5 \pm 1.9$ &  $0.4 \pm 1.0$ &   $1.2 \pm 1.0$ &  $0.5 \pm 1.0$ &  $1.4 \pm 0.7$ &  $1.7 \pm 1.1$ &   $2.9 \pm 0.9$ &   $8.6 \pm 1.1$ &  $16.5 \pm 0.8$ \\
DES16X1eho  &    $2.5 \pm 0.9$ &              - &               - &              - &               - &              - &              - &  $0.0 \pm 0.3$ &   $0.2 \pm 0.4$ &   $0.7 \pm 0.5$ &               - \\
DES17C3gop  &    $2.0 \pm 0.8$ &  $0.2 \pm 0.3$ &   $0.6 \pm 0.3$ &              - &               - &              - &              - &  $0.2 \pm 0.3$ &   $0.6 \pm 0.4$ &   $0.5 \pm 0.2$ &               - \\
DES17S2fee  &    $0.8 \pm 5.1$ &  $0.0 \pm 1.2$ &   $0.0 \pm 1.2$ &  $0.4 \pm 1.0$ &   $1.2 \pm 1.0$ &  $0.4 \pm 1.1$ &  $1.4 \pm 1.0$ &  $0.0 \pm 1.2$ &   $0.6 \pm 1.2$ &   $0.0 \pm 1.0$ &   $2.3 \pm 0.7$ \\
DES17X3dxu  &    $3.1 \pm 0.5$ &              - &               - &              - &               - &              - &              - &  $0.7 \pm 0.5$ &   $0.7 \pm 0.2$ &               - &               - \\
DES17X3cds  &    $3.3 \pm 1.8$ &  $0.4 \pm 0.4$ &   $1.4 \pm 0.4$ &              - &               - &              - &              - &  $0.0 \pm 0.4$ &   $0.3 \pm 0.3$ &   $0.8 \pm 0.3$ &               - \\
DES17C3fwd  &   $14.7 \pm 1.5$ &  $2.0 \pm 0.6$ &   $6.2 \pm 0.6$ &  $0.3 \pm 0.6$ &   $0.8 \pm 0.6$ &              - &              - &  $1.1 \pm 0.7$ &   $0.5 \pm 0.7$ &   $3.8 \pm 0.5$ &   $6.2 \pm 1.5$ \\
DES17X3hxi  &    $5.4 \pm 1.8$ &  $2.3 \pm 0.9$ &   $7.1 \pm 0.9$ &              - &               - &              - &              - &  $1.0 \pm 0.7$ &   $1.6 \pm 0.7$ &   $1.6 \pm 0.5$ &               - \\
DES13E2lpk  &    $8.9 \pm 1.5$ &  $0.4 \pm 0.5$ &   $1.2 \pm 0.5$ &              - &               - &              - &              - &  $0.4 \pm 0.6$ &   $0.8 \pm 0.5$ &   $2.4 \pm 0.4$ &               - \\
DES15C2eal  &    $1.9 \pm 0.6$ &  $0.2 \pm 0.2$ &   $0.6 \pm 0.2$ &  $0.1 \pm 0.2$ &   $0.2 \pm 0.2$ &  $0.4 \pm 0.2$ &  $0.5 \pm 0.2$ &  $0.0 \pm 0.2$ &   $0.3 \pm 0.2$ &   $0.4 \pm 0.3$ &   $2.1 \pm 0.3$ \\
DES16C2ggt  &   $11.8 \pm 1.2$ &  $1.1 \pm 0.4$ &   $3.2 \pm 0.4$ &  $0.4 \pm 0.4$ &   $1.4 \pm 0.4$ &  $1.7 \pm 0.3$ &  $1.1 \pm 0.3$ &  $0.4 \pm 0.5$ &   $1.5 \pm 0.6$ &   $3.4 \pm 0.4$ &   $6.9 \pm 0.5$ \\
DES17C2hno  &    $3.7 \pm 1.1$ &  $0.2 \pm 0.2$ &   $0.7 \pm 0.2$ &              - &               - &              - &              - &  $0.4 \pm 0.5$ &   $0.6 \pm 0.4$ &   $0.9 \pm 0.2$ &               - \\
\bottomrule
\end{tabular}
\end{table*}

\newpage
\section{Bayesian Fits - sSFR}
\label{app:b}

\begin{figure*}
\includegraphics[width=0.5\textwidth]{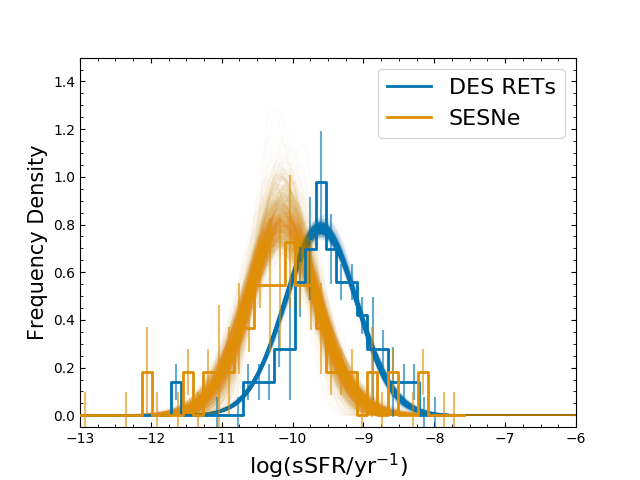}
\caption{Results of the MCMC fits to the PDFs of sSFR for DES RETs and SESNe, accounting for uncertainties in each bin.
\label{fig:histfit_s12_ssfr}}
\end{figure*}

\begin{figure*}
\includegraphics[width=\textwidth]{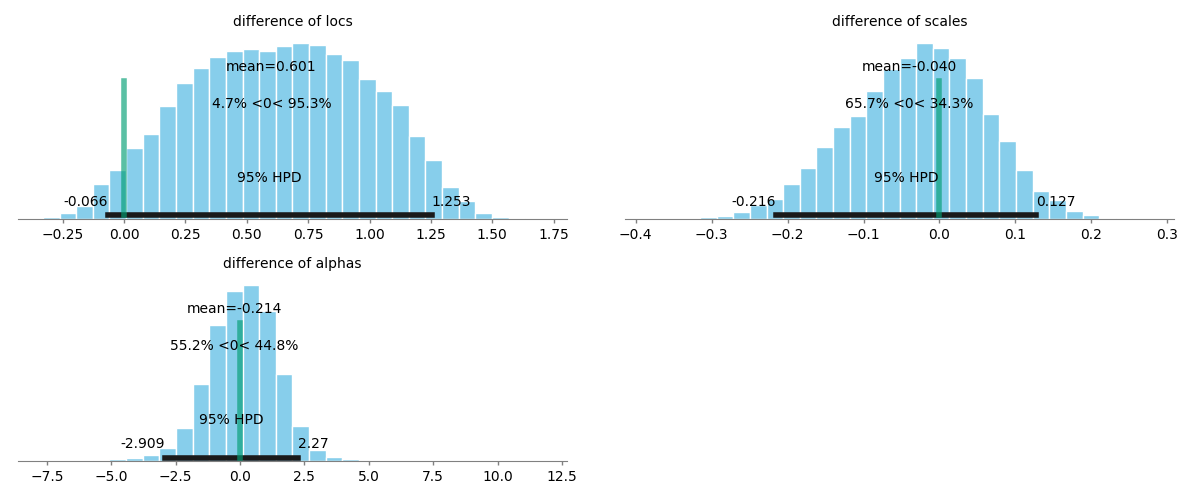}
\caption{Histograms showing the differences between the fit parameters across the MCMC samples for the comparison between RETs and SESNe. 
\label{fig:param_diffs}}
\end{figure*}

\begin{figure*}
\includegraphics[width=\textwidth]{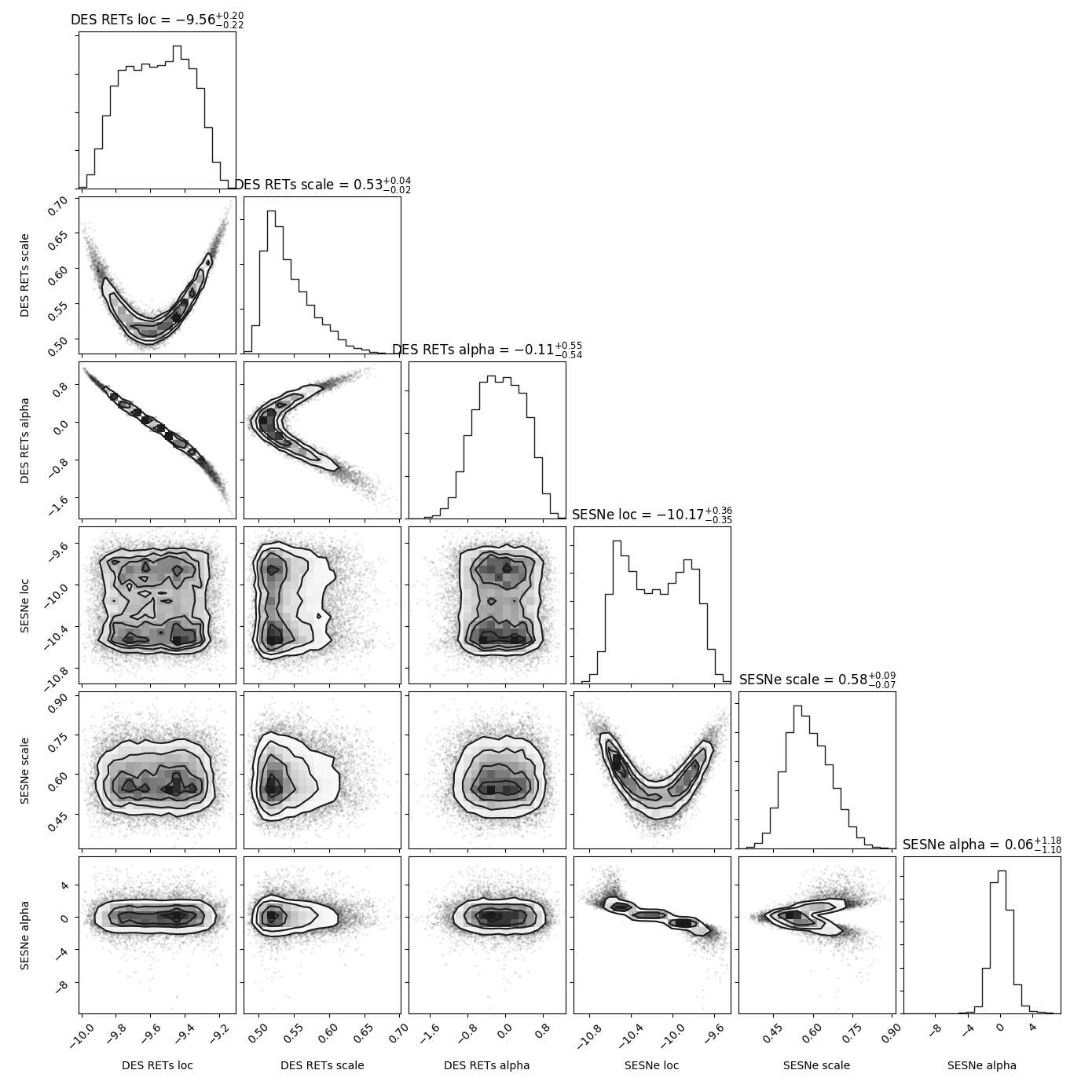}
\caption{Corner plot showing the posterior samples from the MCMC fit to the DES RETs and \citet{Sanders2012} sSFRs. Notable features are: 1) the RET distribution is better constrained than the SESNe (S12); 2) the scale vs alpha and loc vs scale distributions are two-tailed due to alpha being centred close to 0; 3) there is a degeneracy between loc and alpha for the same reason. Loc and scale have units of yr$^{-1}$, while alpha is dimensionless. Figure produced using the \texttt{corner} package \citep{Foreman-Mackey2016}.
\label{fig:corner}}
\end{figure*}

To evaluate the likelihood that two independent distributions are from the same parent population we follow the method outlined in \citetalias{Wiseman2020}. We fit the PDFs, along with the uncertainty on the value in each bin, simultaneously with the same priors using the No U-Turn Sampler (NUTS; \citealt{Hoffman2014}) Hamiltonian Monte Carlo algorithm via the  \texttt{pymc3}\footnote{https://docs.pymc.io/} package to explore the posterior distribution. We utilise two chains, for a warm-up period of $5\times10^3$ iterations per chain and a fit period of $5\times10^3$ iterations per chain. Fig.~\ref{fig:histfit_s12_ssfr} displays an example of the resulting fit where the DES RETs and \citet{Sanders2012} sSFR distributions are compared. Each resulting distribution is described by the `loc' (location), `scale' (spread), and `alpha' (skewness). We then compare the differences in these parameters, as seen in Fig.~\ref{fig:param_diffs}, by reporting objectively the percentages of posterior samples that overlap, and subjectively what this means for the similarity of the distributions.
\twocolumn
\begin{figure}
\includegraphics[width=0.5\textwidth]{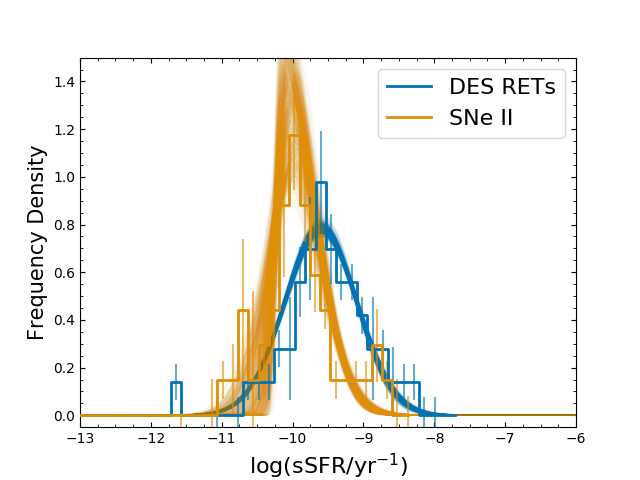}
\caption{Results of the MCMC fits to the PDFs of sSFR for DES RETs and SNe II, accounting for uncertainties in each bin.
\label{fig:histfit_ssfr_s13}}
\end{figure}
\begin{figure}
\includegraphics[width=0.5\textwidth]{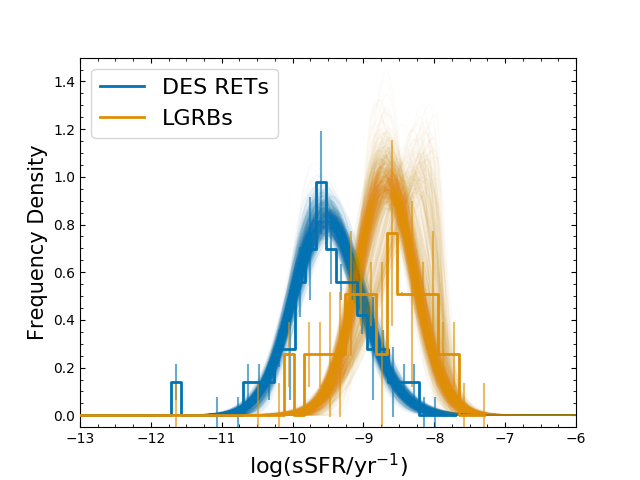}
\caption{Results of the MCMC fits to the PDFs of sSFR for DES RETs and LGRBs, accounting for uncertainties in each bin.
\label{fig:histfit_ssfr_k15}}
\end{figure}
\begin{figure}
\includegraphics[width=0.5\textwidth]{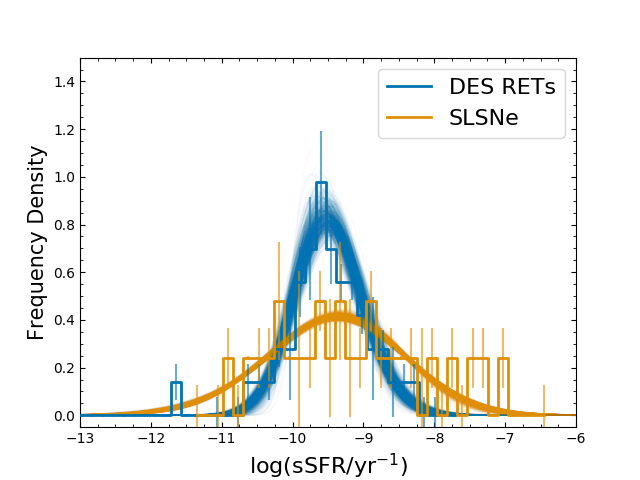}
\caption{Results of the MCMC fits to the PDFs of sSFR for DES RETs and SLSNe, accounting for uncertainties in each bin.
\label{fig:histfit_ssfr_p16}}
\end{figure}

\section{Bayesian Fits - Metallicity}
\label{app:c}
In this section, we present the Bayesian fits to the metallicity distributions of RETs and the comparison samples.
\begin{figure}
\includegraphics[width=0.5\textwidth]{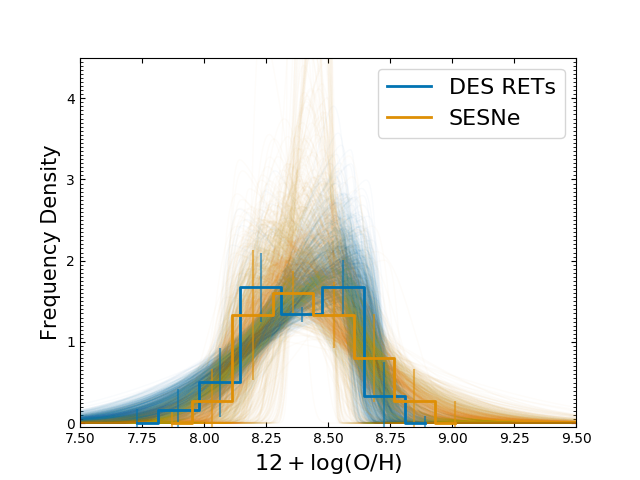}
\caption{Results of the MCMC fits to the PDFs of metallicity for DES RETs and SESNe, accounting for uncertainties in each bin.
\label{fig:histfit_oh_s12}}
\end{figure}

\begin{figure}
\includegraphics[width=0.5\textwidth]{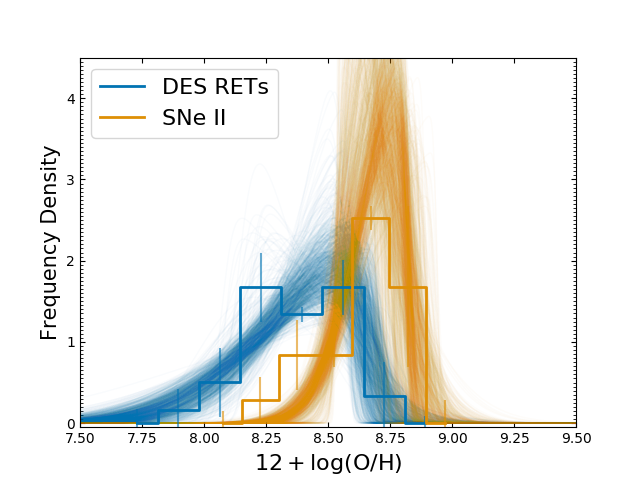}
\caption{Results of the MCMC fits to the PDFs of metallicity for DES RETs and SNe II, accounting for uncertainties in each bin.
\label{fig:histfit_oh_s13}}
\end{figure}

\begin{figure}
\includegraphics[width=0.5\textwidth]{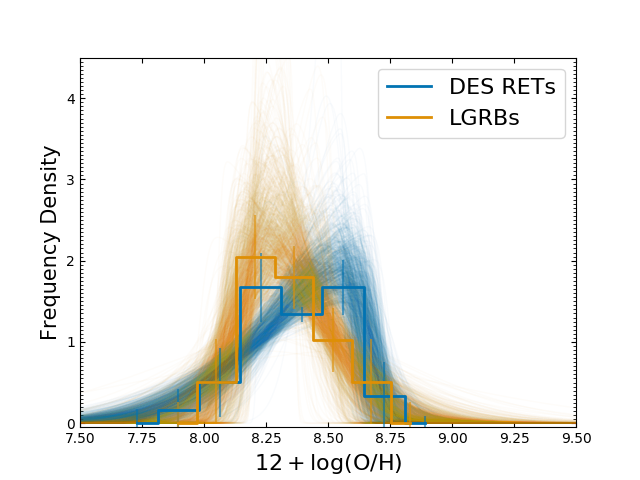}
\caption{Results of the MCMC fits to the PDFs of metallicity for DES RETs and LGRBs, accounting for uncertainties in each bin.
\label{fig:histfit_oh_k15}}
\end{figure}

\begin{figure}
\includegraphics[width=0.5\textwidth]{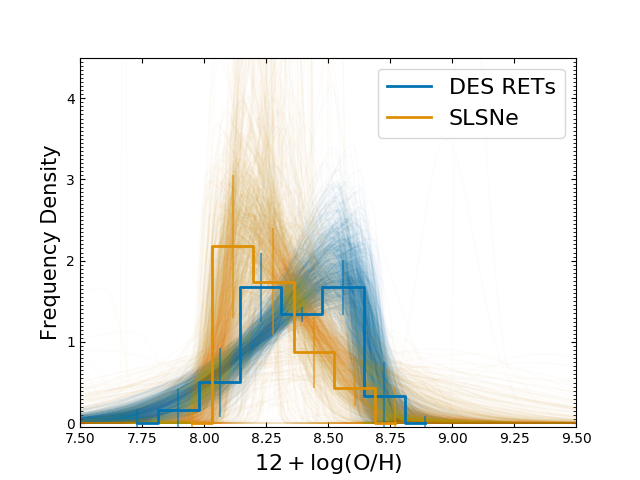}
\caption{Results of the MCMC fits to the PDFs of metallicity for DES RETs and SLSNe, accounting for uncertainties in each bin.
\label{fig:histfit_oh_p16}}
\end{figure}
\onecolumn
\parbox{\textwidth}{
$^{1}$ School of Physics and Astronomy, University of Southampton,  Southampton, SO17 1BJ, UK\\
$^{2}$ Institute of Cosmology and Gravitation, University of Portsmouth, Portsmouth, PO1 3FX, UK\\
$^{3}$ PITT PACC, Department of Physics and Astronomy, University of Pittsburgh, Pittsburgh, PA 15260, USA\\
$^{4}$ The Research School of Astronomy and Astrophysics, Australian National University, ACT 2601, Australia\\
$^{5}$ School of Mathematics and Physics, University of Queensland,  Brisbane, QLD 4072, Australia\\
$^{6}$ Universit\'e Clermont Auvergne, CNRS/IN2P3, LPC, F-63000 Clermont-Ferrand, France\\
$^{7}$ Santa Cruz Institute for Particle Physics, Santa Cruz, CA 95064, USA\\
$^{8}$ Department of Astronomy and Astrophysics, University of Chicago, Chicago, IL 60637, USA\\
$^{9}$ Kavli Institute for Cosmological Physics, University of Chicago, Chicago, IL 60637, USA\\
$^{10}$ Sydney Institute for Astronomy, School of Physics, A28, The University of Sydney, NSW 2006, Australia\\
$^{11}$ Department of Physics and Astronomy, University of Pennsylvania, Philadelphia, PA 19104, USA\\
$^{12}$ Department of Physics, Duke University Durham, NC 27708, USA\\
$^{13}$ Cerro Tololo Inter-American Observatory, National Optical Astronomy Observatory, Casilla 603, La Serena, Chile\\
$^{14}$ Departamento de F\'isica Matem\'atica, Instituto de F\'isica, Universidade de S\~ao Paulo, CP 66318, S\~ao Paulo, SP, 05314-970, Brazil\\
$^{15}$ Laborat\'orio Interinstitucional de e-Astronomia - LIneA, Rua Gal. Jos\'e Cristino 77, Rio de Janeiro, RJ - 20921-400, Brazil\\
$^{16}$ Fermi National Accelerator Laboratory, P. O. Box 500, Batavia, IL 60510, USA\\
$^{17}$ CNRS, UMR 7095, Institut d'Astrophysique de Paris, F-75014, Paris, France\\
$^{18}$ Sorbonne Universit\'es, UPMC Univ Paris 06, UMR 7095, Institut d'Astrophysique de Paris, F-75014, Paris, France\\
$^{19}$ Department of Physics and Astronomy, Pevensey Building, University of Sussex, Brighton, BN1 9QH, UK\\
$^{20}$ Department of Physics \& Astronomy, University College London, Gower Street, London, WC1E 6BT, UK\\
$^{21}$ Kavli Institute for Particle Astrophysics \& Cosmology, P. O. Box 2450, Stanford University, Stanford, CA 94305, USA\\
$^{22}$ SLAC National Accelerator Laboratory, Menlo Park, CA 94025, USA\\
$^{23}$ Centro de Investigaciones Energ\'eticas, Medioambientales y Tecnol\'ogicas (CIEMAT), Madrid, Spain\\
$^{24}$ INAF, Astrophysical Observatory of Turin, I-10025 Pino Torinese, Italy\\
$^{25}$ Department of Astronomy, University of Illinois at Urbana-Champaign, 1002 W. Green Street, Urbana, IL 61801, USA\\
$^{26}$ National Center for Supercomputing Applications, 1205 West Clark St., Urbana, IL 61801, USA\\
$^{27}$ Institut de F\'{\i}sica d'Altes Energies (IFAE), The Barcelona Institute of Science and Technology, Campus UAB, 08193 Bellaterra (Barcelona) Spain\\
$^{28}$ INAF-Osservatorio Astronomico di Trieste, via G. B. Tiepolo 11, I-34143 Trieste, Italy\\
$^{29}$ Institute for Fundamental Physics of the Universe, Via Beirut 2, 34014 Trieste, Italy\\
$^{30}$ Observat\'orio Nacional, Rua Gal. Jos\'e Cristino 77, Rio de Janeiro, RJ - 20921-400, Brazil\\
$^{31}$ Institut d'Estudis Espacials de Catalunya (IEEC), 08034 Barcelona, Spain\\
$^{32}$ Institute of Space Sciences (ICE, CSIC),  Campus UAB, Carrer de Can Magrans, s/n,  08193 Barcelona, Spain\\
$^{33}$ Instituto de Fisica Teorica UAM/CSIC, Universidad Autonoma de Madrid, 28049 Madrid, Spain\\
$^{34}$ Centre for Astrophysics \& Supercomputing, Swinburne University of Technology, Victoria 3122, Australia\\
$^{35}$ Department of Physics, Stanford University, 382 Via Pueblo Mall, Stanford, CA 94305, USA\\
$^{36}$ Center for Cosmology and Astro-Particle Physics, The Ohio State University, Columbus, OH 43210, USA\\
$^{37}$ Department of Physics, The Ohio State University, Columbus, OH 43210, USA\\
$^{38}$ Center for Astrophysics $\vert$ Harvard \& Smithsonian, 60 Garden Street, Cambridge, MA 02138, USA\\
$^{39}$ Australian Astronomical Optics, Macquarie University, North Ryde, NSW 2113, Australia\\
$^{40}$ Lowell Observatory, 1400 Mars Hill Rd, Flagstaff, AZ 86001, USA\\
$^{41}$ George P. and Cynthia Woods Mitchell Institute for Fundamental Physics and Astronomy, and Department of Physics and Astronomy, Texas A\&M University, College Station, TX 77843,  USA\\
$^{42}$ Department of Astronomy, The Ohio State University, Columbus, OH 43210, USA\\
$^{43}$ Instituci\'o Catalana de Recerca i Estudis Avan\c{c}ats, E-08010 Barcelona, Spain\\
$^{44}$ Institute of Astronomy, University of Cambridge, Madingley Road, Cambridge CB3 0HA, UK\\
$^{45}$ Department of Astrophysical Sciences, Princeton University, Peyton Hall, Princeton, NJ 08544, USA\\
$^{46}$ Department of Physics, University of Michigan, Ann Arbor, MI 48109, USA\\
$^{47}$ Computer Science and Mathematics Division, Oak Ridge National Laboratory, Oak Ridge, TN 37831\\
$^{48}$ Max Planck Institute for Extraterrestrial Physics, Giessenbachstrasse, 85748 Garching, Germany\\
$^{49}$ Universit\"ats-Sternwarte, Fakult\"at f\"ur Physik, Ludwig-Maximilians Universit\"at M\"unchen, Scheinerstr. 1, 81679 M\"unchen, Germany\\
}

\bsp	
\label{lastpage}
\end{document}